\documentclass[preprint2]{aastex}
\usepackage{natbib}
\usepackage{graphicx,graphics,subfigure}

\bibpunct{(}{)}{;}{a}{}{,}

\slugcomment{To appear in the Astronomical Journal}

\shorttitle{{\em Spitzer} observations of A1763: I}
\shortauthors{Edwards et al.}

\begin{document}

\title{{\em Spitzer} observations of Abell 1763 - II: Constraining the nature of activity in the cluster-feeding filament with VLA and XMM-Newton data}

\author{Louise O. V. Edwards and Dario Fadda
\affil{NASA Herschel Science Center, Caltech 100-22, Pasadena, CA 91125}}
\email{louise@ipac.caltech.edu}
\author{David T. Frayer
\affil{NRAO, Green Bank, WV, 24944}}
\author{Gastao B. Lima Neto
\affil{Universidade de Sao Paulo, Inst Astronomico e Geofisico, Cidade Universitaria, BR Sao Paulo, SP 05508-900, Brazil}}
\author{Florence Durret
\affil{Institut d'Astrophysique de Paris, 98bis Bd Arago, FR75014, Paris, France}}

\begin{abstract}

The Abell 1763 superstructure at z=0.23 contains the first galaxy filament to be directly detected using mid-infrared observations. Our previous work has shown that the frequency of starbursting galaxies, as characterized by 24$\mu$m emission is much higher within the filament than at either the center of the rich galaxy cluster, or the field surrounding the system. New VLA and XMM-{\it Newton} data are presented here. We use the radio and X-ray data to examine the fraction and location of active galaxies, both active galactic nuclei (AGN) and starbursts. The radio far-infrared correlation, X-ray point source location, IRAC colors, and quasar positions are all used to gain an understanding of the presence of dominant AGN. We find very few MIPS-selected galaxies that are clearly dominated by AGN activity. Most radio selected members within the filament are starbursts. Within the supercluster, 3 of 8 spectroscopic members detected both in the radio and in the mid-infrared are radio-bright AGN. They are found at or near the core of Abell~1763. The five starbursts are located further along the filament. We calculate the physical properties of the known wide angle tail (WAT) source which is the brightest cluster galaxy (BCG) of Abell~1763. A second double lobe source is found along the filament well outside of the virial radius of either cluster.  The velocity offset of the WAT from the X-ray centroid, and the bend of the WAT in the intracluster medium (ICM) are both consistent with ram pressure stripping, indicative of streaming motions along the direction of the filament. We consider this as further evidence of the cluster-feeding nature of the galaxy filament.

\end{abstract}


\section{Introduction}

Most of the galaxies in massive clusters have long since been thought to be old, red, and retired. But, galaxy clusters themselves are dynamic systems. Hierarchical structure formation models show these systems to be connected by large scale filaments \citep{bon96,gon09} as they build up over time via mergers and interactions of smaller systems \citep{nav95,kau99,spr05}.  Many local clusters appear dynamically relaxed with Gaussian velocity distributions \citep{mil05,fin08}, smooth X-ray temperature and luminosity profiles \citep{per98,whi00,vik05}, and large central galaxies at their gravitational potential well \citep{jon84,del07}. On the other hand, tell-tail signs of ongoing and recent cluster activity is likewise seen: chaotic mass profiles \citep{zha09,ume09}, central galaxies with wide-angle tailed (WAT) morphology \citep{sak00,smo07}, a displacement between the central galaxy and cluster mass centroid \citep{clo06,coz09}, and a neighboring concentration of quasars \citep{soc04} are some examples.

Abell 1763 is one such dynamic cluster. We have gathered multi-wavelength photometry of the galaxies of this supercluster - Abell~1763, its poorer companion Abell~1770, and the large-scale cluster feeding galaxy filament which connects the two. An excess of infra-red (IR) bright galaxies within the filament was announced in \nocite{fad08} Fadda et al. 2008 (hereafter Paper~0). Our catalog of optical, Near-IR, Mid-IR and Far-IR infrared photometry is presented in Edwards et al. 2010 (hereafter Paper~I).\nocite{edw10}

An ambiguity exists in the possible origins of the bright mid-IR detected galaxies. Most of this emission is expected to arise from current star formation, as UV light from hot stars suffers extinction through dust in the galaxy. However, emission from obscured AGN can also cause significant IR radiation. 

In dust obscured galaxies, the optical emission lines can lead to erroneous measurements of star formation rates (SFRs) without a solid knowledge of the extinction \citep{kew02}. Also, optical surveys can miss up to 50\% of an AGN population in luminous infrared galaxies \citep{ima07}. Similarly it has long been known that radio AGN do not always harbor optical emission lines \citep{owe95,bes05}.  Fortunately, the radio emission at 1.4$\,$GHz is well known to correlate with the MIPS 24$\mu$m and 70$\mu$m  emission from starbursts \citep{hel85,con92,yun01}. The radio emission is the synchrotron emission of relativistic electrons from past supernovae and, to some extent, free-free emission in HII regions associated with young ionizing stars. Sources with stronger radio emission relative to the FIR-radio starburst correlation can be attributed to the overpowering effect of an intense AGN. This provides a method for selecting galaxies clearly dominated by radio-excess AGN. Considering that the FIR star formation rates rely on the presence of dust, the radio SFRs provide a natural complement. The redder Mid-IR continuum and lack of PAH emission in AGN can also be pinpointed by examining the IRAC colors \citep{eis04,lac04}.

A wide-angle tail radio galaxy lives at the center of Abell~1763. The suggestion has been that the peculiar shape of WAT galaxies is from ram pressure caused by relative motions between the cluster gas and the plumes of the radio galaxy \citep{owe76}. Most WATs are found to be strongly associated with optical brightest cluster galaxies (BCGs) \citep{bur81,har04}, as is the case for Abell~1763. However, strong ram pressure resulting from streaming motions of the BCG orbiting the cluster potential at high speeds seems unlikely as the galaxy should be near the bottom of the potential well of the cluster \citep{mil72}. \citet{sak00}, \citet{har04}, and \citet{smo07} suggest that the ram pressure resulting from the displacement of BCGs after interactions with sub-clusters is adequate in creating the WAT morphology. 

Just as the small scale structure of the WAT hints at the cluster kinematics, so does the large scale structure. To probe this, the level of substructure and merging history of the cluster can be gleaned from the X-ray surface brightness distribution. 

In this paper we investigate the AGN and starburst (SB) activity in the different environments of the superstructure and further argue for an ongoing merger and flow of filament galaxies into Abell~1763. Observations in the radio, optical, MIR and X-ray do not always yield the same type of AGN \citep{hic09,gri10} and so we quantify the AGN and SB sources integrating the FIR-radio relation, X-ray point sources, and IRAC colors. We also find an overdensity of quasars near the cluster region. We separate the frequency of AGN and SB activity in the cluster versus the filament. We consider the special case of the WAT of Abell~1763, its morphology and location. Finally, we use unpublished archival XMM observations and describe the WAT and filament orientation with respect to the cluster X-ray structure.  To achieve these goals, {\it Spitzer} photometry from Paper~I is used in conjunction with new wide field 1.4$\,$GHz VLA radio observations which cover Abell~1763 and the filament galaxies towards Abell~1770, and new XMM data which cover the cluster core. 

Data and observations are discussed in Section~2, in Section~3 we show our results: an inventory of AGN in the supercluster, radio SFRs, and a physical description of the central WAT, and the radio SFRs.  In Section~4 we discuss relative contributions of AGN and SBs, as well as the galaxy activity as a function of environment. The interplay between the large scale (X-ray profile) and small scale (WAT) structures lead us to conclude in Section~5 that the filament galaxies are dominated by SB systems and enhance our argument of a cluster feeding filament. Throughout the paper we adopt a cosmology with $\Omega$$_{m}$=0.3, $\Omega$$_{\lambda}$=0.7 and H$_{o}$=74km/s/Mpc where 1$\,$arcsec=3.5$\,$kpc at a redshift of 0.23, unless otherwise stated.

\section{Observations and Data Reduction}

The analysis is based in part on {\it Spitzer} MIPS 24$\mu$m and 70$\mu$m observations discussed in Paper~I, as well as optical spectra to be published in a forthcoming paper (Fadda et al, in progress). Additionally, we provide results from new radio 1.4GHz observations across the Abell~1763-Abell~1770 supercluster from the VLA and XMM-{\it Newton} X-ray observations of the cluster core.

\subsection{VLA Observations and Reduction}

\begin{figure*}
\epsscale{2.3}
\plotone{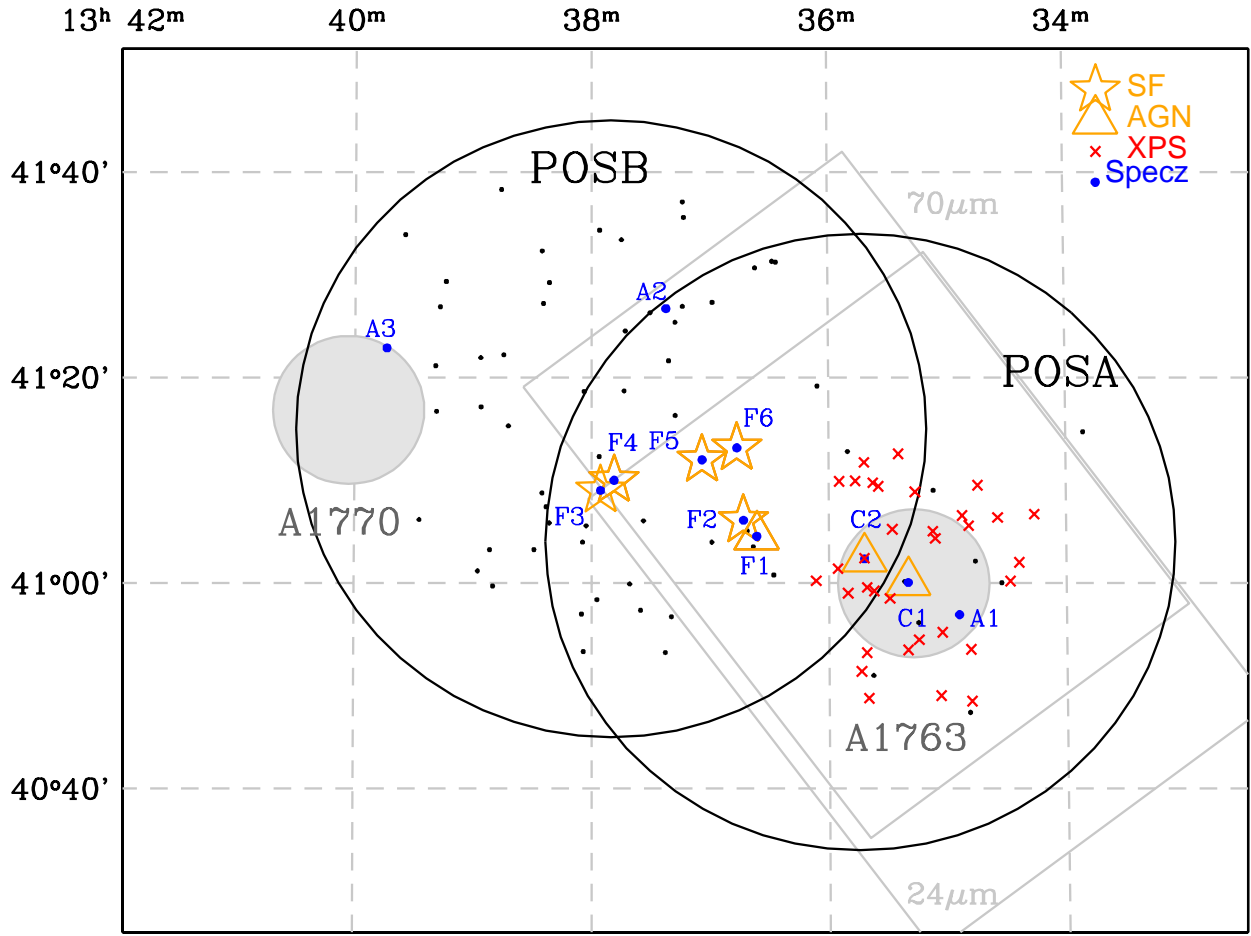}
\caption{{\bf Coverage of photometric and spectroscopic data.} Large black circles trace the FOV of the two 1.4GHz pointings and the relative cluster positions, represented by large filled grey circles. Three $\sigma$ radio detections are indicated by black dots. Those which are spectroscopically confirmed members are filled in dark blue. The 8 member radio sources with MIPS$\,$24$\mu$m or 70$\mu$m detections are labeled with orange. The MIPS coverage is also shown as rectangles. XMM observations cover the core of Abell~1763 and X-ray point sources are marked as small red crosses. All galaxies detected in the radio and FIR lie in the cluster or towards the galaxy filament, the five furthest along the filament are star forming. Cluster core members are labeled `C', those in the filament `F', and those with absorption line spectra `A'.}\label{radcov}
\end{figure*}

We mapped the supercluster using two pointings illustrated in Figure~\ref{radcov} as large black circles. Cluster core members are labeled `C', those in the filament `F', and those with absorption line spectra `A'. Both were made in the continuum at 1.4$\,$GHz in 25$\,$MHz bandwidth spectral line mode with the VLA B-configuration. A spatial resolution of 5$\,$arcsec was achieved which is of order that of the MIPS 24$\mu$m observations of Paper~I.  The observations were carried out in 30 blocks of 1.0$\,$hr over 16 nights between March 27 and April 13th, 2009 as listed in Table~\ref{obs}. Here we also list the primary beam at the half power point (PB). The calibrator source 1331+305 was observed at the beginning and end of each block. It was bright, stable, and close to the target cluster so was used as both the complex gain and bandpass calibrator.

\begin{deluxetable}{lcccccc}
\tabletypesize{\scriptsize}
\tablewidth{0pt}
%\tablecolumns{<num>}
\tablecaption{Radio Observations \label{obs}}
\tablehead{\colhead{Pointing} & \colhead{RA (J2000)} & \colhead{DEC (J2000)} & \colhead
{N blocks} & \colhead{IT/block(min)}& \colhead{PB(arcmin)} &\colhead
{RMS ($\mu$Jy)}}
\startdata
A1763A &  13 35 45.00 & +41 04 00.000 & 30 &     20.68 & 30 & 132 \\
A1763B &  13 37 50.00 & +41 14 00.000 & 30 &     20.68 & 30 & 28\\
1331+305 & 13 31 08.29 & +30 30 32.958 & 30 &     16.38 & - & - \\
\enddata
\end{deluxetable}

%macro read RadioPOS_agn.sm

The data reduction and visualization was performed using AIPS. We followed the procedures outlined in the AIPS cookbook~\footnote{http://www.aips.nrao.edu/CookHTML/CookBook.html} as well as for the particular case of wide field L-band imaging as outlined on the NRAO webpages~\footnote{http://www.vla.nrao.edu/astro/guides/lowfreq/analysis/} and in \citet{owe08}. A first round of flagging was performed on the uncalibrated $u$-$v$ data using UVFLAG and TVFLAG to remove dysfunctional antennas and bad time periods. The data taken on different days were immediately combined using DBCON and INDXR. Spurious $u$-$v$ data points with uncalibrated fluxes greater than 6$\,$Jy for the calibrator, and greater than 1$\,$Jy for the sources in A1763A (the central pointing) and in A1763B (the pointing towards the filament), were also removed.

The calibrator was then SPLIT from the rest of the data, and phase self-calibrated using CALIB. We used BPASS to produce the bandpass correction table which is applied directly to the other sources. 

The 1st-pass calibration used observations of the calibrator, 1331+305. The flux of the primary calibrator was found using SETJY and VLACALIB created the SN table (the solution table stores the complex gains determined from the calibrator). We use antenna 19 for reference as it is near the center of the array and reliable during the entire observation period. VLACLCAL was then run creating the CL table to be applied to the data. This calibration table stores smoothed complex gains and corrections as a function of time to be applied to each source.

Because there are sufficiently bright sources in the pointings, self-calibration was used to refine the accuracy. Here we split the two pointings and self-calibrated each separately. Boxes were set around all real sources, and for each pointing, IMAGR and CALIB were used iteratively. First phase only calibrations are twice applied, then phase+amplitude. For A1763A, the BCG is very bright and PEELR was run to self-calibrate separately from the rest of the field. Finally, PBCOR was run on each pointing to correct for the primary beam. Reduced data from the different stokes and IFs were averaged together. An RMS map for each pointing was created using RMSD.

\subsection{XMM Observations and Reduction}

The Observation Data Files (ODF ID 0084230901) of Abell~1763 were retrieved from the XMM-\textit{Newton} archive. We then processed them using the standard tasks \textit{emchain} and \textit{epchain} from the Science Analysis System (SAS) version 8.0.0. The observation was done in full window extended with the medium filter.

We produced ``clean'' event files by filtering out events with pattern larger than 12 in MOS and 4 in pn, and selecting only events in the FOV. The light curves in the [10--12 keV] band showed that there were high particle background time intervals during the observations. Filtering out the periods with flares reduced the exposure times from 17.4~ks to 13.6~ks (MOS) and from 11.6~ks to 9.2~ks (pn).

With the cleaned and filtered event files, we created the redistribution matrix file (RMF) and ancillary response file (ARF) with the SAS tasks \texttt{rmfgen} and \texttt{arfgen} for each camera and for each region analyzed.

The background was taken into account by extracting spectra from the blank sky templates described by \citet{Carter07} and reprojected to the coordinates and roll angle of the Abell~1763 observation. The background spectra for each detector were normalized following the procedure used in \citet{Lagana08}, by comparing each to the background extracted from the cluster observation itself, but in an annulus between 12.5$\,$ and 14$\,$arcmin, where contribution from Abell~1763 is not detected.

\subsection{Radiometry and Catalog Matching}

We used the AIPS source detection algorithm, SAD, to derive fluxes for radio sources down to the 3$\sigma$ level in both fields. The BCG, an extended source, has a total flux of 350$\,$mJy, and unfortunately lies close to the center of the A1763A pointing, limiting the dynamic range. The measured RMS near the center is therefore 132$\,$$\mu$$\,$Jy (and 2$\times$132$\,$$\mu$$\,$Jy at the half power point), whereas it is only 28$\,$$\mu$Jy (2$\times$28$\,$$\mu$$\,$Jy at the half power point) in the center of the A1763B pointing. In order to analyze comparable detection limits between the two pointings the 3$\sigma$ sources in A1763A should be compared to the 14$\sigma$ sources in A1763B (a flux limit of $\sim$0.40$\,$mJy in both cases).

SAD fails for very large and extended sources. We therefore visually examined the catalog. In cases where it was clear that a large source has been broken up into several smaller sources, we removed these false detections from the catalog and instead included the result from IMEAN, which produce the peak and integrated flux densities in a region specified with TVWIN and give similar results to those obtained from JMFIT. We provide the RMS error given by IMEAN but note it is similar to the RMS value from the map created with RMSD within the same area the photometry was measured in. In the few cases with a clear lobe+central source morphology, we kept the object as one detection and included the flux from the lobes as well as the central source. 

\begin{deluxetable}{cccccccccc}
\tabletypesize{\scriptsize}
\tablewidth{0pt}
%\tablecolumns{<num>}
\tablecaption{Radio Fluxes for Cluster Members\label{clrads}}
\tablehead{\colhead{RA (J2000)} & \colhead{DEC (J2000)} & \colhead
{Peak Flux} & \colhead{Total Flux}& \colhead{Pointing} &\colhead{Photo z} &\colhead{Spec z} & \colhead{q$_{24}$} & \colhead{q$_{70}$} & \colhead{Label}\\ \colhead{(deg)} & \colhead{(deg)} & \colhead
{(mJy/bm)} & \colhead{(mJy)}& \colhead{} &\colhead{} &\colhead{} &\colhead{} &\colhead{} & \colhead{}}
\startdata
203.729874 & 40.607719 &   49.34 $\pm$    3.43 &  130.54 $\pm$   12.01 & A1763A & 0.2560 & - & -1.90 &    - &\\ 
204.127914 & 40.944340 &   23.78 $\pm$    0.13 &   26.04 $\pm$    0.24 & A1763A & 0.2472 & - & -2.07 &    - &\\ 
203.722885 & 40.945854 &   18.29 $\pm$    0.31 &   68.59 $\pm$    1.42 & A1763A & 0.1930 & - & -1.47 &    - &\\ 
203.833618 & 41.001053 &  205.10 $\pm$    0.13 &  349.90 $\pm$    0.32 & A1763A & 0.2334 & 0.2280 & -3.24 &    - & C1\\ 
204.321533 & 41.034107 &    0.18 $\pm$    0.06 &    1.23 $\pm$    0.43 & A1763B & 0.2669 & - &  0.55 &  1.59 &\\ 
203.925201 & 41.039402 &    0.92 $\pm$    0.28 &    0.98 $\pm$    0.51 & A1763A & 0.1806 & 0.2181 &  0.23 &    -& C2\\ 
204.152267 & 41.076385 &    2.07 $\pm$    0.03 &    6.03 $\pm$    0.10 & A1763A & 0.1949 & 0.2345 & -0.18 &  0.57 & F1 \\ 
204.364975 & 41.082771 &    0.07 $\pm$    0.05 &    0.21 $\pm$    0.19 & A1763B & 0.2311 & - &  0.72 &    - &\\ 
204.179306 & 41.101704 &    0.22 $\pm$    0.03 &    0.19 $\pm$    0.04 & A1763A & 0.1563 & 0.2316 &  1.73 &  2.40& F2\\ 
204.480927 & 41.150002 &    0.29 $\pm$    0.05 &    0.28 $\pm$    0.08 & A1763B & 0.1866 & 0.2311 &  1.17 &    - & F3\\ 
204.452316 & 41.166481 &    0.42 $\pm$    0.04 &    0.37 $\pm$    0.07 & A1763B & 0.2277 & 0.2564 &  0.66 &  1.82& F4\\ 
204.265793 & 41.199951 &    0.28 $\pm$    0.05 &    1.41 $\pm$    0.30 & A1763B & 0.2529 & 0.2576 &  0.81 &  2.14& F5\\ 
204.194122 & 41.219238 &    0.22 $\pm$    0.06 &    0.66 $\pm$    0.24 & A1763B & 0.1229 & 0.2330 &  0.57 &  2.07 & F6\\ 
204.547623 & 41.312935 &    1.02 $\pm$    0.04 &    1.13 $\pm$    0.09 & A1763B & 0.1999 & - &    - &  1.22& \\ 
204.475113 & 41.288013 &    0.09 $\pm$    0.04 &    0.07 $\pm$    0.06 & A1763B & 0.2031 & - &    - &  2.09& \\ 
\hline
203.726074 & 40.948601 &    2.05 $\pm$    0.34 &    1.71 $\pm$    0.52 & A1763A & 0.2458 & 0.2380 & - & - & A1\\ 
203.841034 & 41.002407 &    7.84 $\pm$    0.37 &   10.19 $\pm$    0.76 & A1763A & 0.2604 & - & - & -& \\ 
204.304626 & 41.592812 &    1.55 $\pm$    0.25 &   10.39 $\pm$    1.90 & A1763B & 0.2391 & - & - & -& \\ 
204.342300 & 41.444931 &    1.35 $\pm$    0.11 &    3.18 $\pm$    0.36 & A1763B & 0.2263 & 0.2564 & - & - & A2\\ 
204.690735 & 41.638103 &    1.80 $\pm$    0.32 &   22.38 $\pm$    4.30 & A1763B & 0.1817 & - & - & - &\\ 
204.932495 & 41.381260 &    1.12 $\pm$    0.03 &    2.39 $\pm$    0.08 & A1763B & 0.2700 & 0.2535 & - & - & A3\\ 
204.515198 & 41.310566 &    0.46 $\pm$    0.04 &    0.46 $\pm$    0.08 & A1763B & 0.1839 & - & - & - &\\ 
204.675247 & 41.254757 &    0.22 $\pm$    0.04 &    0.63 $\pm$    0.17 & A1763B & 0.2667 & - & - & - &\\ 
204.491089 & 41.281853 &    0.12 $\pm$    0.04 &    0.42 $\pm$    0.16 & A1763B & 0.2200 & - & - & - &\\ 
204.549469 & 41.160805 &    0.11 $\pm$    0.04 &    0.79 $\pm$    0.36 & A1763B & 0.2660 & - & - & - &\\ 
204.613235 & 41.226471 &    0.11 $\pm$    0.04 &    0.44 $\pm$    0.20 & A1763B & 0.2257 & - & - & - &\\ 
204.638367 & 41.196552 &    0.15 $\pm$    0.04 &    0.23 $\pm$    0.10 & A1763B & 0.2259 & - & - & - &\\ 
204.685532 & 41.100777 &    0.18 $\pm$    0.05 &    0.91 $\pm$    0.30 & A1763B & 0.1994 & - & - & - &\\ 
204.780914 & 41.262924 &    0.17 $\pm$    0.06 &    0.70 $\pm$    0.32 & A1763B & 0.2745 & - & - & - &\\ 

\enddata
\end{deluxetable}
%OLD awk '($24<0.26&&$24>0.21)||($22<0.28&&$22>0.18){printf "%10.6f & %9.6f & %7.2f & %7.2f & %7.2f & %7.2f & %d & %4.1f & %6.4f & %6.4f \\\\ \n", $3,$4,$5/1000,$6/1000,$7/1000,$8/1000,$9,$21,$22,$24}' A1763_Radio_SM.ASC
%FOR RADIO+MIPS awk '$25!=396&&$25!=84&&$25>=0&&(($16>=0.21&&$16<=0.26)||(($14<=0.28&&$14>=0.18)&&($16<=0))){printf "%10.6f & %9.6f & %7.2f \\pm %7.2f & %7.2f \\pm %7.2f & %7.2f  %7.2f & %6.4f & %6.4f & %6.3f & %6.3f \\\\ \n", $2,$3,$25/1000,$26/1000,$27/1000,$28/1000,$23,$24,$14,$16,log($17/$25)/log(10),log($19/$25)/log(10)}' test2470_NoRad.ASC
%FOR RADIO_NOMIPS awk '(($24<0.26&&$24>0.21)||(($22<0.28&&$22>0.18)&&$24<=0))&&$25<=0&&$27<=0&&$21<=3.0{printf "%10.6f & %9.6f & %7.2f \\pm %7.2f & %7.2f \\pm %7.2f & %d & %6.4f & %6.4f & %1s & %1s \\\\ \n", $3,$4,$5/1000,$6/1000,$7/1000,$8/1000,$9,$22,$24,"-","-"}' A1763_Radio_SM.ASC

To discriminate the radio sources between members and non-members of the superstructure, we matched the $\sim$614 3$\sigma$ radio detections to optical and redshift catalogs. The Sloan Digital Sky Survey, DR7 (hereafter SDSS) covers the entire field in the u$^{\prime}$,g$^{\prime}$,r$^{\prime}$,i$^{\prime}$, and z$^{\prime}$ bands. In this paper, the optical colors, all photometric redshifts and many spectroscopic redshifts are collected from the SDSS. In addition, we have obtained many spectra from runs with the Hydra instrument on WIYN and from TNG (Fadda et al. 2010, in progress). We use the reliability method of \citet{cil03} and \citet{sut92} to compute the most probable optical association for the radio observation. This method takes into account not only the distance from both detections, but also the probability of matching to a background object by computing the overdensity of optical sources as a function of magnitude. We compute reliabilities from all possible optical associations $<$ 3.5$\,$arcsec of the radio detection in magnitude bins of 0.5 from r$^{\prime}$=15-25. There are 135 matches, 117 of which have very good reliabilities of $>$0.90. All have a reliability $>$0.44. Of the 135 radio sources that have optical associations, 30 have spectra (with 13 confirmed cluster members) and an additional 21 have photometric redshifts between 0.18 and 0.28. Certainly some of the photometric members are real, but for this analysis we will focus only on the spectroscopically confirmed cluster members.

%awk '($19>0.21&&$19<0.26)||($19<0&&$17>0.18&&$17<0.28){print $0}' A1763_RadioSDSSR.ASC | wc -l

Table~\ref{clrads} shows the radio flux densities for possible cluster members with those that have confirmed redshifts labeled. The first two columns give the Right Ascension and Declination. Columns 3 and 4 list the peak flux per beam and total flux for the 5$\sigma$ detections (we include the additional 3$\sigma$ sources in A1763B). The errors quoted with the fluxes are computed assuming the central RMS value for photometry done with SAD. For the sources whose photometry was measured using IMEAN, we measured the local RMS value. The 5$\sigma$ error for pointing A1763A is 0.5$\,$mJy and 0.2$\,$mJy for pointing A1763B, which does not include the absolute flux calibration error of $\sim$15\%. The total flux is always less than the peak flux, within the stated errors. The next column gives the pointing. The error on the VLA positions is $\sim$0.6~FWHM/(SNR) \citep{con97}, which is typically $\sim$1$\,$arcsec for our data. We include only the 28 matches which have counterparts closer than 3.0$\,$arcsec.  The photometric and spectroscopic redshifts are listed in the next two columns. The values for q$_{24}$ and q$_{70}$ are included where available for the spectroscopic sources, and the galaxy label is included in the last column.

The $\sim$1 degree region around Abell 1763 which covers most of A1763A and approximately half the length of the filament has been observed with {\it Spitzer}.  MIPS 24$\mu$m counterparts to the radio sources were found by using the same reliability method we used to choose the SDSS counterparts. The MIPS 70$\mu$m counterpart is taken to be that closest to the radio source, and must be within 10$\,$arcsec.

\begin{figure}
 \subfigure{
     \begin{minipage}[c]{0.45\textwidth}
        \centering
        \includegraphics[width=\textwidth]{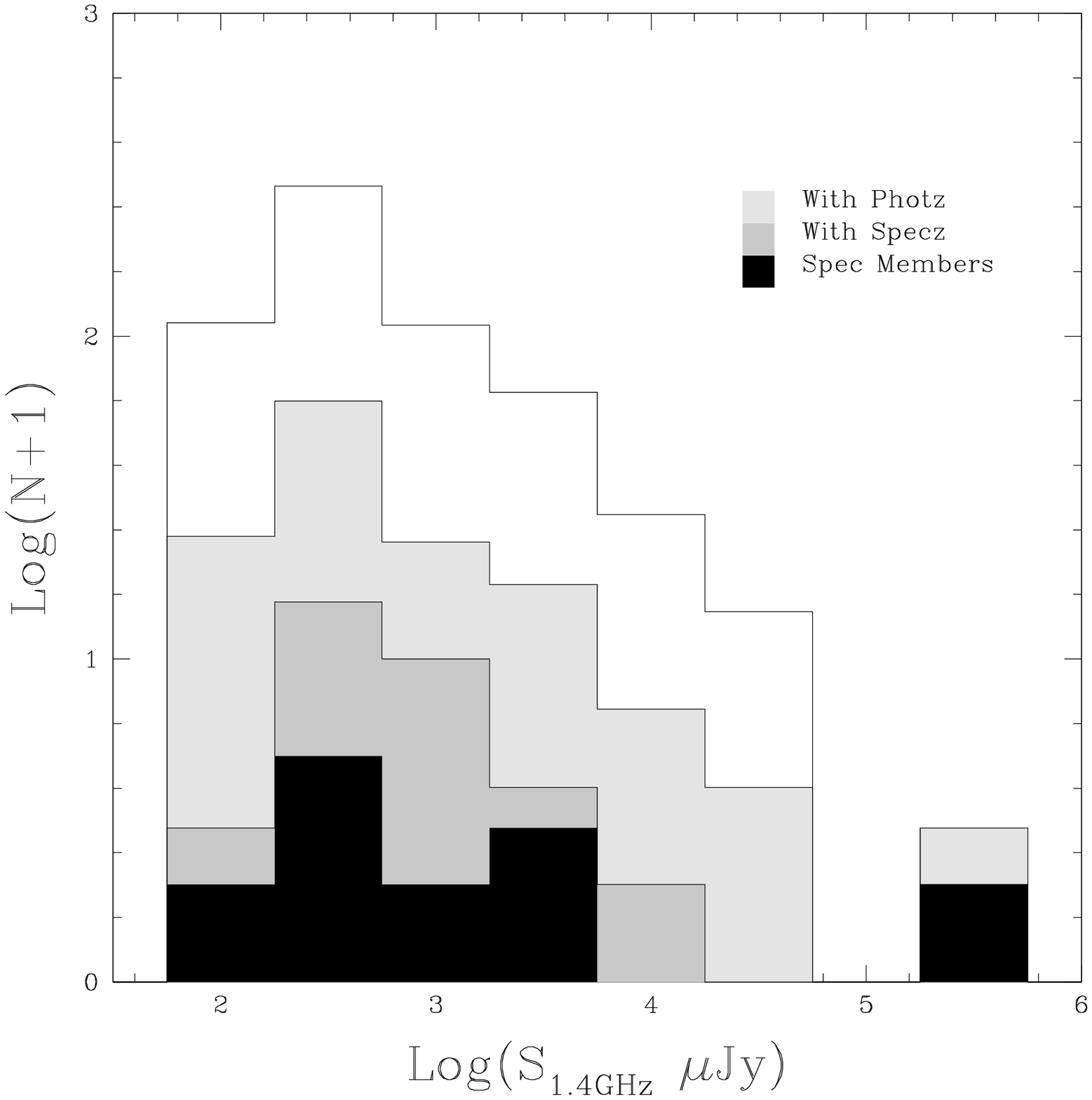}
         \end{minipage}%
    }
 \subfigure{
 \begin{minipage}[c]{0.45\textwidth}
        \centering
        \includegraphics[width=\textwidth]{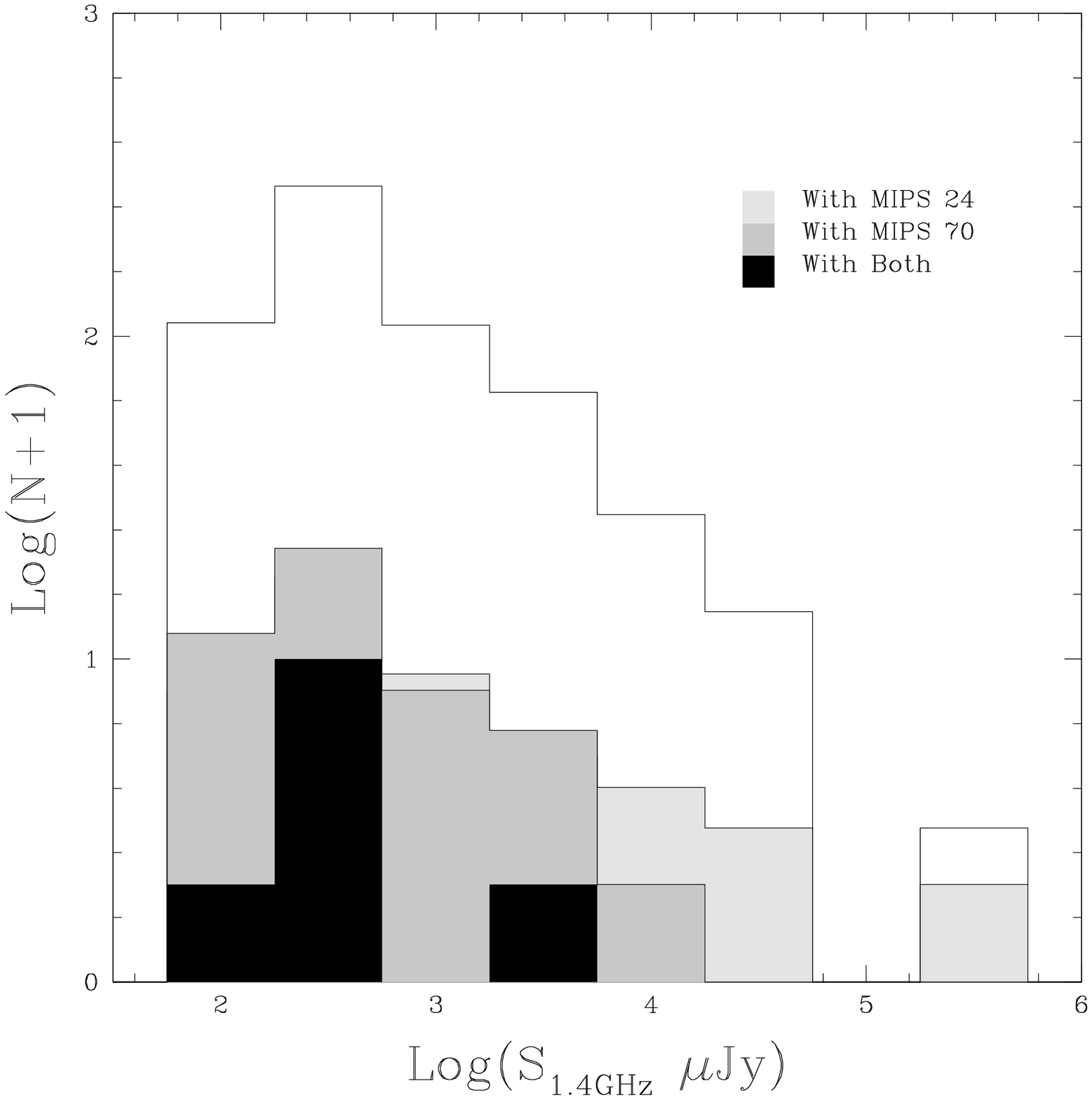}
        \end{minipage}%
    }
    \caption[]{{\bf Histograms of the radio source flux density}. Top: galaxies with photometric redshifts are in light grey, with spectroscopic redshifts are in dark grey, and spectroscopically confirmed supercluster members are in black. Most radio sources do not have bright optical companions and hence the number of redshifts is limited. Bottom: light grey represents galaxies with 24$\mu$m detections, dark grey represents those with 70$\mu$m detections, and black those with detections at both wavelengths. The shape of the distribution for the full sample is echoed in the FIR-detected subsamples. \label{radhist}}
\end{figure}
% SM  macro read radioHist.sm radioHistMip.sm

Figure~\ref{radhist} shows a histogram of radio detected source fluxes from both the A1763A and A1763B fields. The top plot shows 614 3$\sigma$ radio sources in the region, 135 have SDSS detections, and 30 have spectroscopic counterparts. Thirteen of the radio-selected sources have spectroscopic redshifts between 0.21 and 0.26 (the redshift of Abell 1763 is 0.23). Most radio sources have flux densities between 500-1000$\,\mu$Jy and the distribution of spectroscopic members is the same as for the entire population. The histogram on the bottom shows that 76 have MIPS 24$\mu$m or 70$\mu$m counterparts. The distribution of radio fluxes for the MIPS sub-samples echos that of the full sample.

\subsubsection{Detection Limits}

\begin{figure}
 \subfigure{
     \begin{minipage}[c]{0.45\textwidth}
        \centering
        \includegraphics[width=\textwidth]{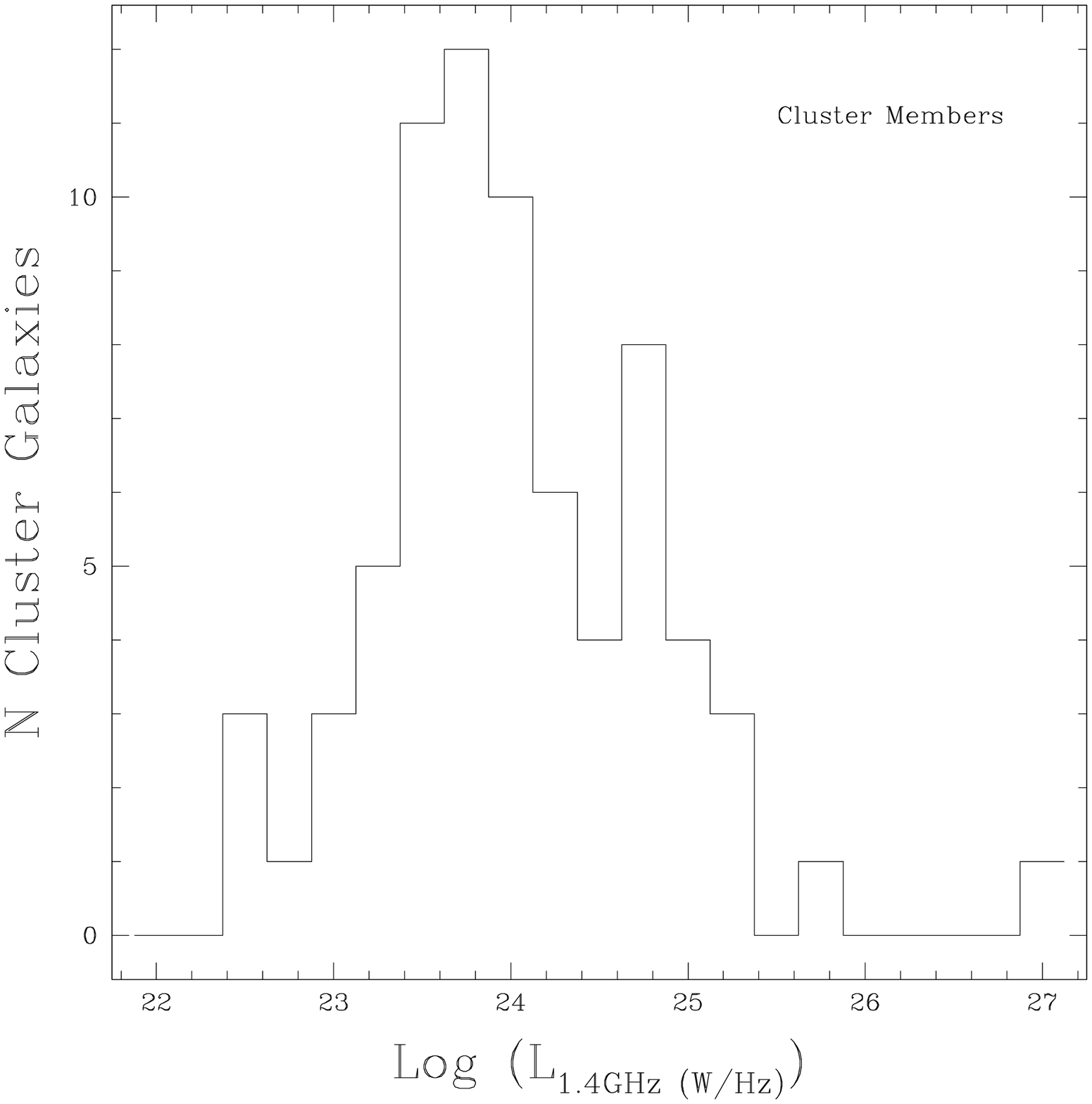}
         \end{minipage}%
    }
 \subfigure{
 \begin{minipage}[c]{0.45\textwidth}
        \centering
        \includegraphics[width=\textwidth]{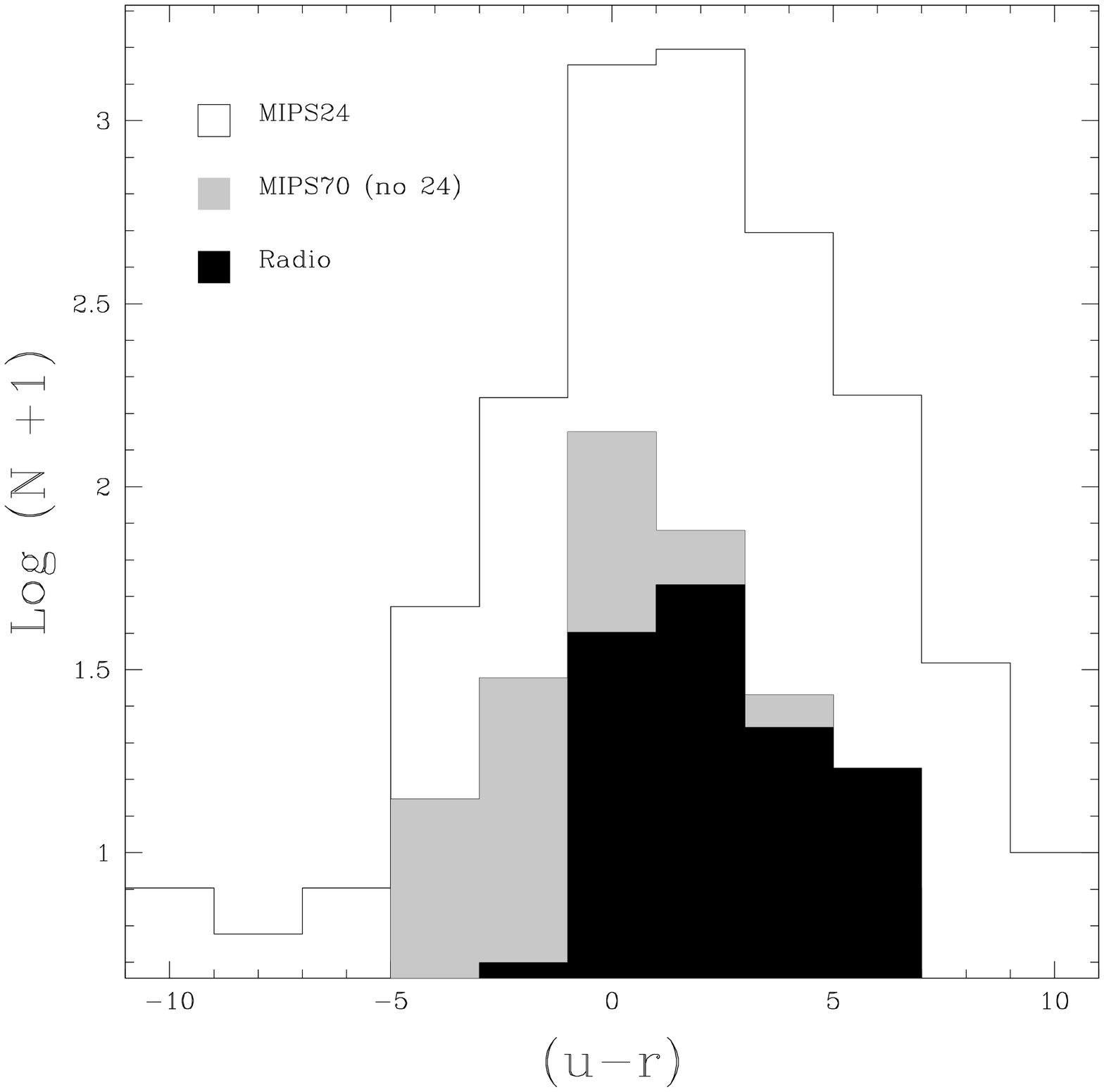}
        \end{minipage}%
    }
    \caption[]{{\bf Histograms of the radio source luminosity and of optical colors.}
Top: a histogram of the radio source luminosity for only the galaxies in the cluster, where galaxies have photometric redshifts between 0.18 and 0.28 or spectroscopic redshifts between 0.21 and 0.26. Bottom: The white histogram shows the optical colors of all MIPS 24 detections (members and non-members) which also have SDSS counterparts. The grey histogram plots the colors of those with MIPS 70 detections, and the black histogram shows colors for the galaxies with radio detections as well. The radio galaxies display a range of radio intensity and those MIPS sources with radio and optical counterparts show the same distribution of optical colors. \label{lumhist}}
\end{figure}

% HistRLum.sm from RadProp.f and HistRM.sm
For the radio sources with spectra, we can calculate the limiting radio luminosity at 1.4$\,$GHz. The top panel of Figure~\ref{lumhist} shows the histogram of radio luminosities for galaxies at the redshift of the cluster. The bottom panel shows the optical colors for all galaxies (members and non-members) located in the region observed by both MIPS and the VLA. At the mean redshift of the cluster, z=0.23, the distance is 1085$\,$Mpc. The limiting 3$\sigma$ 1.4$\,$GHz flux density of 0.4$\,$mJy in A1763A leads to log~L$_{1.4GHz}$~=~22.75, log~L$_{1.4GHz}$~=~22.08 for A1763B). Notice that the distribution of colors is similar for the population of MIPS galaxies as a whole (white histograms), as for the subset of radio-detected galaxies (black histograms). However, the radio sample does miss the bluest galaxies.

Within our detection limits, we are sensitive to all the radio loud and most radio excess galaxies. \citet{yun01} define radio-loud galaxies as those with 1.4$\,$GHz luminosities greater than $\sim$10$^{25}$ W/Hz. We find 2 of these in the velocity range of the cluster. Radio excess, on the other hand, are defined as having radio emission 5 times more than expected from the radio-FIR correlation (q$_{24}$$<$0.3, where q$_{24}$ is defined as log(S$_{24 \mu m}$/S$_{1.4GHz}$)). The MIPS 24$\,\mu$m 3$\sigma$ limiting flux corresponds to a minimum radio flux of 40$\,\mu$Jy for radio-excess AGN.

\subsection{X-ray analysis}

Fitting a one component MEKAL model to the XMM data using \textsc{xspec} version 11.3.2, the mean emission weighted temperature, and metallicity for the intracluster gas was obtained inside a circle of 3.5$\,$arcmin radius ($735 \,$kpc). With the hydrogen column density fixed at the Galactic value $0.82 \times 10^{20}\,$cm$^{-2}$ (using the \textit{nh} tool from \textit{ftools}, based on the Leiden/Argentine/Bonn Survey) we obtained the temperature $kT = 6.8 \pm 0.3$~keV and metallicity $Z = 0.29 \pm 0.06 Z_{\odot}$ (both values at 90\% confidence level).

Inside the 3.5$\,$arcmin radius, the unabsorbed X-ray flux in the [0.5--10.0 keV] band is $f_{X} = (7.8 \pm 0.3) \time 10^{-12}\,$erg~s$^{-1}$~cm$^{-2}$. The corresponding luminosity is $L_{X} = (1.2 \pm 0.1) \times  10^{45}\,$erg~s$^{-1}$. The unabsorbed bolometric X-ray luminosity (assuming an interval 0.01--100.0 keV) is $L_{X,bol} = (1.7 \pm 0.1) \times 10^{45}\,$erg~s$^{-1}$.

\section{Results}

Here we report three main results. We show that the AGN count is low, calculate the radio SFRs for the probable SBs, and derive the WAT properties. 

\subsection{Most FIR sources are starbursts}

To help distinguish likely AGN from SBs for the cluster and filament MIPS and radio selected galaxies, we use various complementary methods. We check the radio-FIR correlation, find X-ray point sources, compare IRAC colors, and finally cross-correlate the positions with those of known quasars.

\subsubsection{The radio-FIR correlation}

  \begin{figure*}
   \center
  \epsscale{2.0}
     \plotone{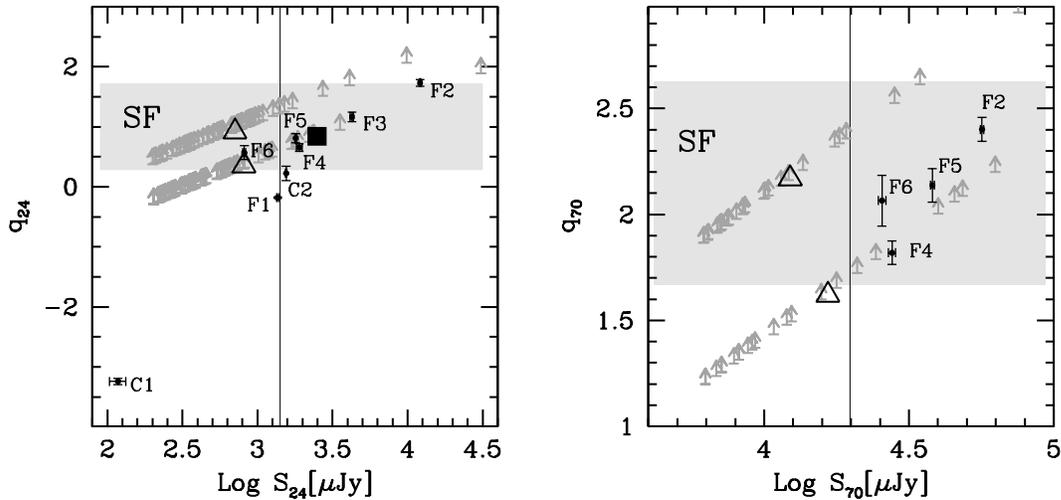}
    \caption{{\bf The radio-FIR correlation.} The plots show the logarithm of the observed flux density ratio (q) versus the infrared flux density for all spectroscopically confirmed cluster sources with a MIPS detection. Sources with MIPS 24$\mu$m emission to the left, and with MIPS 70 $\mu$m emission are to the right. Sources with non-detections in the radio are marked as grey upper limits, with the average value shown as large open triangles. The solid grey bar indicates the best-fit radio-FIR correlation and its 3$\sigma$ error, q$_{24}$$\sim$ 1$\,\pm\,$0.72 and q$_{70}$$\sim$ 2.15$\,\pm\,$0.48. The detection limit of position A1763A is marked as a vertical black line. The region under the correlation band is expected to be populated only by radio-excess AGN. The black square marks the q-value for the MIPS-faint stack discussed in Section~\ref{stacksec}. 
\label{qval}}
\end{figure*}

To determine the presence of radio-excess AGN we show the radio-FIR correlation for star forming galaxies \citep{con92} in Figure~\ref{qval}. Sources well below the relation, with a radio excess are usually attributed to AGN emission. From \citet{app04}, q$_{24}$$\sim$ 1$\,\pm\,$0.24 and q$_{70}$= log~(S$_{70\mu m}$/S$_{1.4GHz}$) $\sim$ 2.15$\,\pm\,$0.16. In our figure and analysis we consider the 3$\sigma$ error on the q-value instead of the 1$\sigma$ levels published in \citet{app04}. A note of caution: we are limited by the 3$\sigma$ radio detection limits as the MIPS data is deeper than the radio data. 

  \begin{figure*}
   \center
  \epsscale{2.0}
     \plottwo{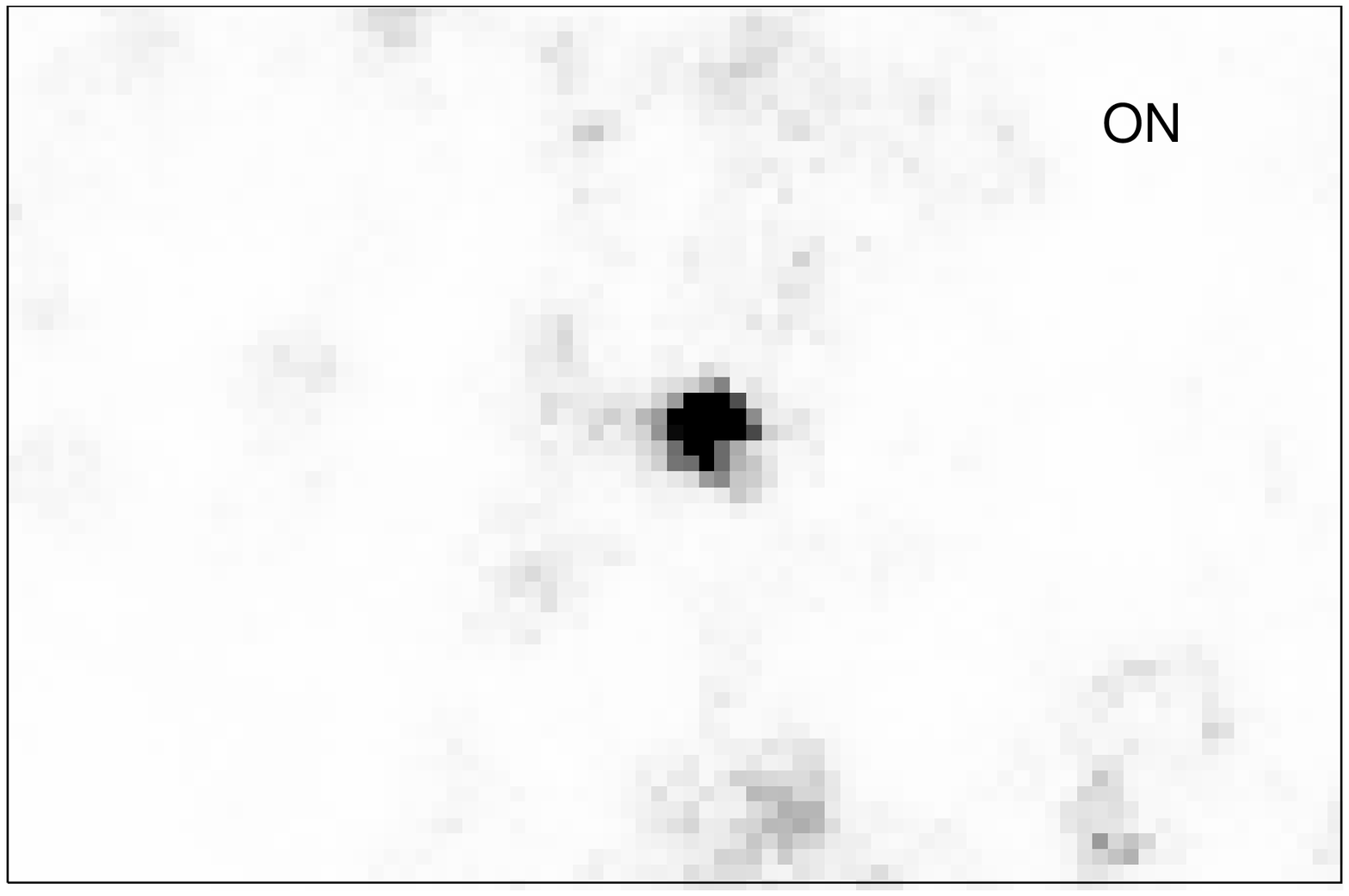}{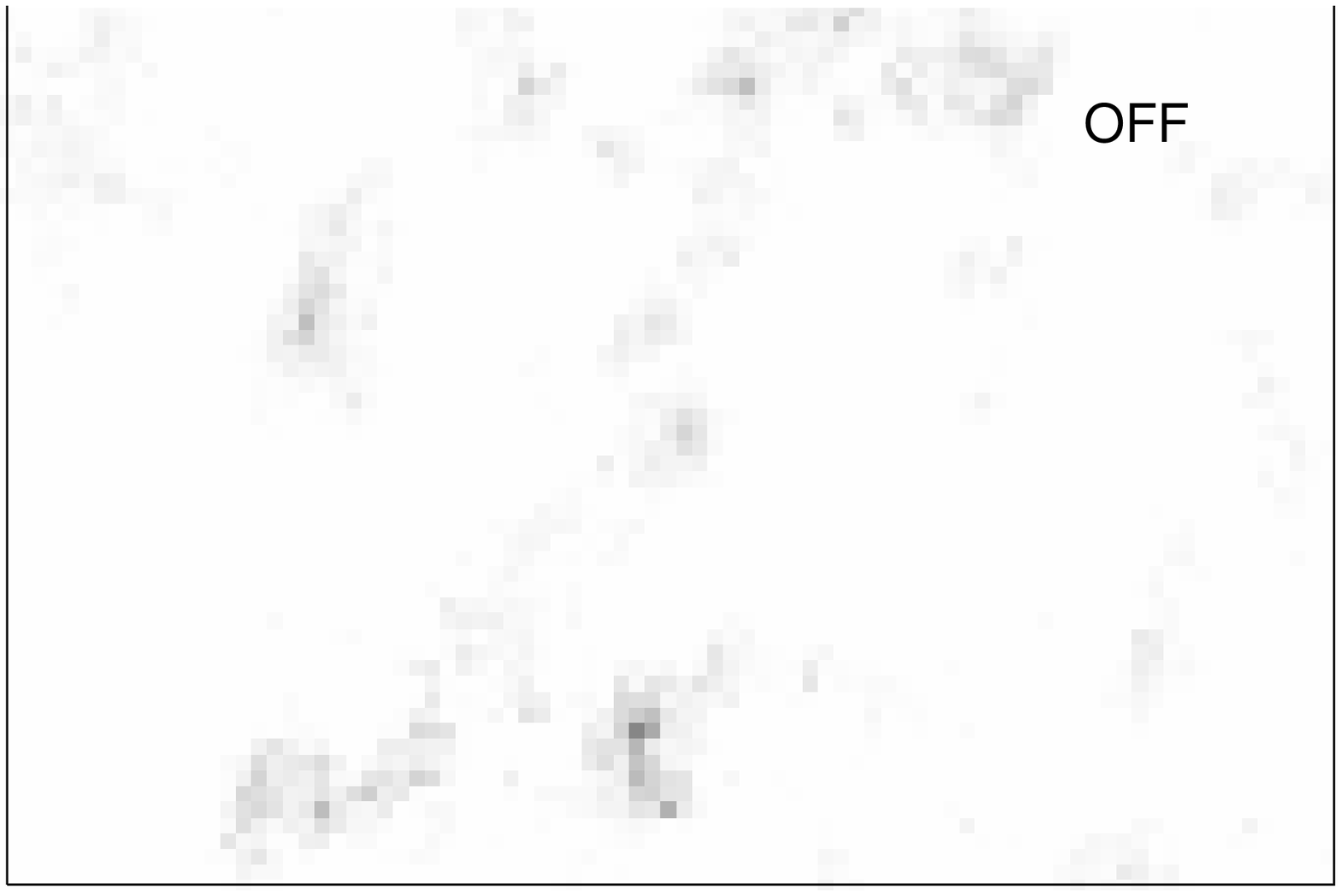}
    \caption{{\bf The faint MIPS 24$\mu$m radio stack.} A stacking analysis of the faint sources confirms that the average population of faint MIPS galaxies is consistent with SB activity. The area shown is $\sim$2.8$\,$arcmin  across. On the right is a stack of off-MIPS positions.
\label{mstack}}
\end{figure*}

Most of the MIPS sources from Paper~I are not detected in the Radio and we have upper limits for most of these sources. They correspond to lower limits on the q-values of Figure~\ref{qval}. All cluster galaxies falling in the deeper Radio pointing are classified as SB. Unfortunately, the faint MIPS end in pointing A1763A can not all be confirmed SB from their upper limits. We calculate the average values for the upper limits in both pointings to find that both are indeed within the SB portion of the diagnostic (open triangle of Figure~\ref{qval}). We measure the q$_{24}$ value for 8 radio bright member galaxies. Five are classified SB galaxies, all along the filament as shown in Figure~\ref{radcov}. F1 is an AGN within the filament and C1 and C2 are AGN inside the core of Abell~1763. All of the galaxies classified as SB through q$_{24}$ are likewise classified as SB through q$_{70}$.

\label{stacksec}
For the MIPS sources that are so faint that AGN and SB radio emission would both be below our radio detection limit (MIPS 24$\mu$m flux $<$400$\mu$Jy), we performed a stacking analysis. We perform this for the A1763B field since it has a deeper signal. There are 626 faint MIPS sources with high signal to noise ratios ($>$10) and MIPS $<$ 4$\,$mJy, and the median MIPS flux is 2.5$\,$mJy. We create individual postage stamps in the radio surrounding the central pixel of the MIPS source which we then smooth by the PSF of the MIPS data. The postage stamps are median combined together and a faint signal is detected as shown in Figure~\ref{mstack}. The average radio flux is 36$\,$$\mu$Jy with an RMS of 2.5$\,$$\mu$Jy (over 1000 pixels). Therefore, the q$_{24}$$\sim$0.84$\pm$0.04, is well within the radio-FIR correlation as shown in Figure~\ref{qval} as the filled square. Therefore at least the average q$_{24}$ for the faint MIPS population is consistent with a population of star forming galaxies. Only 142 of the above are confirmed cluster members (we lower our allowed SNR to $>$3). Repeating the same analysis, the radio flux is smaller with only 19$\,$$\mu$Jy (RMS=4.6$\,$$\mu$Jy over 1320 pixels) and q$_{24}$ is even higher at 1.26 (-1.14 +1.36).

\subsubsection{X-ray point sources}

We have selected all XMM point sources that have more than 25 raw counts in the 0.5--8.0 keV band in the XMM-pn observation, corresponding approximately to $10^{-14}\,$erg~s$^{-1}$~cm$^{-2}$, for this observation. To ensure the AGN nature of the XMM point sources, we estimate the hardness ratio, defined as\\
$$ \mbox{HR} =  \frac{\mbox{hard} - \mbox{soft}}{\mbox{hard} + \mbox{soft}}\, ,$$ \\
 where ``soft'' is the soft band 0.5--1.5~keV and ``hard'' is the hard band 1.5--8.0~keV. With the hardness ratio, we used \textit{xspec} to estimate the spectrum slope, assuming the point source emission could be described by a power-law spectrum with only the Galactic absorption (i.e., no in-situ absorption).

The above procedure revealed that the brightest point source in the Abell~1763 field is probably either a nearby Young Star Object (YSO) or an evolved giant red star (as seen by a bright optical counterpart on SDSS). In both cases, the X-ray emission, which is very soft, is due to wind collision. There is another source with very soft emission without an obvious optical counterpart in SDSS. The other point sources are all probably bona fide AGN, given their slope around 1.8. The point sources that are projected onto the cluster core are listed in Table~\ref{psprop}. If the point sources have detections either in at 24$\mu$m or 1.4$\,$GHz, this is noted in the table.

There are other XMM point sources, for which we do not have spectra that may, or may not be part of the superstructure. However, only one additional source has a MIPS detection, therefore it will not affect our statistics.

The point sources with measured redshifts within the superstructure velocity include only one radio detected source, at position C2 of Figure~\ref{radcov}. We note that there is also one bright ROSAT point source outside our XMM field of view and along the filament, at position F3 of Figure~\ref{radcov}. 

\begin{deluxetable}{lcccccc}
\tabletypesize{\scriptsize}
\tablewidth{0pt}
%\tablecolumns{<num>}
\tablecaption{XMM Point Sources in the core of A1763\label{psprop}}
\tablehead{\colhead{RA (J2000)} & \colhead{DEC (J2000)} & \colhead{Flux} &\colhead{HR} &\colhead{indxPL}&\colhead{errIndxPL} &\colhead{Notes}\\\colhead{deg} & \colhead{deg} & \colhead{10$^{-14}$$\,$erg/s/cm$^{2}$} &\colhead{} &\colhead{}&\colhead{}&\colhead{}}
\startdata
203.781342 &  41.083797 & 20.0 & -0.38 & 2.94 & 0.13 & 24$\mu$m \\ 
203.924561 &  41.039997 & 10.0 & 0.76 & 0.94 & 0.10 & 24$\mu$m 1.4GHz, C2\\ 
203.811462 &  40.907810 & 7.3 & 0.39 & 1.73 & 0.16 &  \\ 
203.834412 &  40.891369 & 5.6 & 0.38 & 1.76 & 0.19 &  \\ 
203.872421 &  40.974945 & 5.0 & 0.55 & 1.46 & 0.18 & 24$\mu$m \\ 
203.960388 &  40.983299 & 3.9 & 0.59 & 1.36 & 0.20 & 24$\mu$m \\ 
203.701523 &  40.892181 & 3.8 & 0.01 & 2.32 & 0.26 &  \\ 
203.853317 &  41.209587 & 3.7 & 0.76 & 0.95 & 0.17 & 24$\mu$m 1.4GHz\\ 
203.919540 &  40.992699 & 3.6 & 0.61 & 1.33 & 0.20 &  \\ 
203.977859 &  41.164612 & 3.5 & 0.70 & 1.10 & 0.19 & 24$\mu$m \\ 
203.762054 &  40.920235 & 3.2 & 0.49 & 1.55 & 0.23 &  \\ 
203.866241 &  41.086914 & 3.2 & 0.52 & 1.50 & 0.23 &  \\ 
203.643890 &  41.106319 & 2.8 & 0.47 & 1.60 & 0.25 & 24$\mu$m \\ 
203.719391 &  41.109531 & 2.8 & 0.67 & 1.18 & 0.22 &  \\ 
203.895355 &  41.156796 & 2.6 & 0.39 & 1.74 & 0.27 & 24$\mu$m \\ 
203.598740 &  41.033463 & 2.5 & 0.88 & 0.42 & 0.14 &  \\ 
204.026993 &  41.003651 & 2.4 & 0.23 & 1.99 & 0.30 &  \\ 
203.917191 &  40.813114 & 2.3 & 0.50 & 1.55 & 0.27 &  \\ 
203.924774 &  41.195778 & 2.0 & 0.76 & 0.95 & 0.23 &  \\ 
203.980515 &  41.023273 & 2.0 & 0.63 & 1.27 & 0.26 &  \\ 
203.905045 &  40.986900 & 2.0 & 0.59 & 1.37 & 0.28 &  \\ 
203.944214 &  41.165539 & 2.0 & 0.57 & 1.40 & 0.28 &  \\ 
203.764511 &  40.817234 & 1.9 & 0.33 & 1.84 & 0.33 &  \\ 
203.921417 &  40.887112 & 1.9 & 0.49 & 1.56 & 0.30 &  \\ 
203.567184 &  41.111725 & 1.8 & 0.48 & 1.58 & 0.32 & 24$\mu$m \\ 
203.705261 &  41.093075 & 1.7 & 0.61 & 1.33 & 0.29 &  \\ 
203.906876 &  41.162567 & 1.7 & 0.72 & 1.04 & 0.26 &  \\ 
203.931854 &  40.856575 & 1.5 & 0.57 & 1.41 & 0.32 &  \\ 
203.776382 &  41.072380 & 1.5 & 0.79 & 0.84 & 0.25 &  \\ 
203.617737 &  41.003082 & 1.3 & 0.59 & 1.37 & 0.34 &  \\ 
203.818329 &  41.147701 & 1.1 & -0.45 & 3.09 & 0.55 &  \\ 
203.701401 &  40.808292 & 1.1 & 0.80 & 0.79 & 0.28 &  \\ 
203.686234 &  41.158386 & 1.0 & 0.18 & 2.06 & 0.47 &  \\ 
\enddata
\end{deluxetable}

\subsubsection{AGN from the IRAC-IRAC colors}

\begin{figure}
\epsscale{1.3}
\plotone{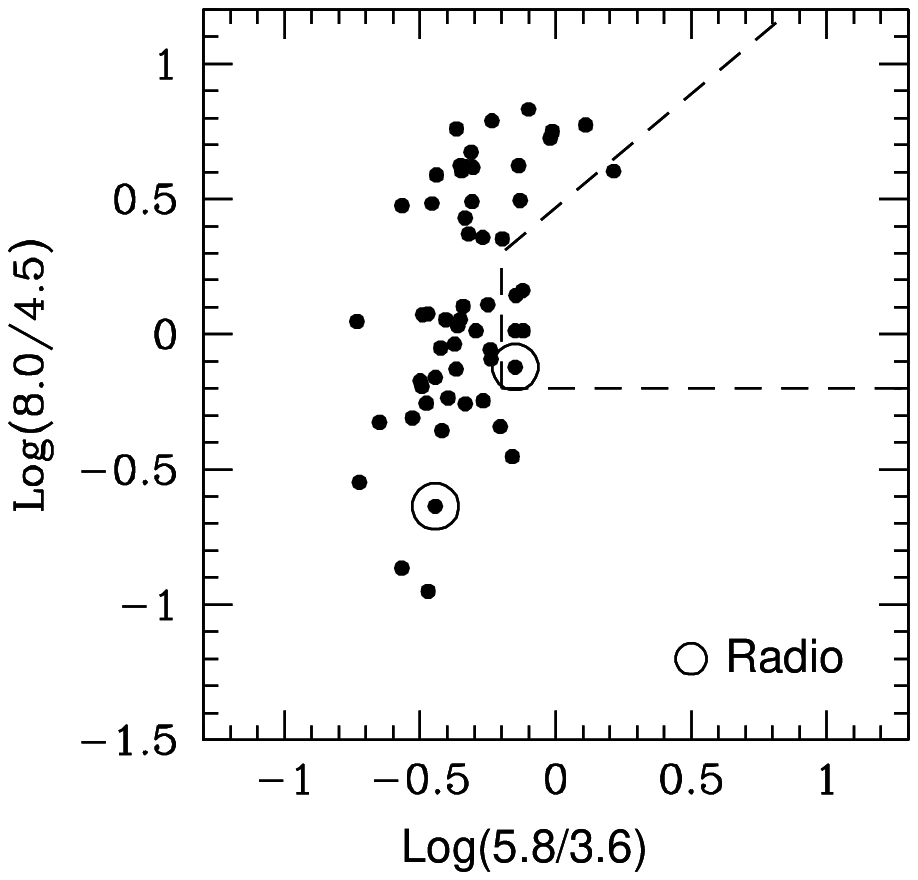}
\caption{{\bf IRAC Colors.} Diagnostic diagram of \citet{lac04} for the MIPS selected galaxies which have detections in all four IRAC bands. Filled black circles are spectroscopic members. Very few of the spectroscopically confirmed members are inside of the dashed polygon, the AGN region of the diagnostic. The two galaxies in the sample with radio detections are F1 and C1. }\label{iracAGN}
\end{figure}

IRAC colors can be used as diagnostics to separate stars, galaxies, and broad line AGN \citep{eis04,lac04,saj05,don08}. AGN are well separated because the power law AGN emission is much redder that the galaxy spectrum in the first filter, and the lack of PAHs in AGN differentiates them from the galaxies in the second filter. \citet{lac04} plot 8.0/4.5 vs 5.8/3.6 which separates very well the objects with blue continua from objects with red continua, a good diagnostic of an AGN is if red colors are seen in both IRAC filters. 

We plot the IRAC colors of our MIPS selected galaxies from Paper~I in Figure~\ref{iracAGN}. Very few of the spectroscopically confirmed members fall into the AGN region of the \citet{lac04} diagnostic diagram. This is consistent with the model results of \citet{saj05} and template results of \citet{don08} which both find that for low z galaxies, the \citet{lac04} diagnostic plot does a good job of separating AGN from the SB sample. Of our five galaxies that fall into the AGN region of the \citet{lac04} diagnostic (Table~\ref{iracagntab}), only one source, the radio-excess AGN F1 of Figure~\ref{radcov}, is also a radio source.

\begin{deluxetable}{lcccccc}
\tabletypesize{\scriptsize}
\tablewidth{0pt}
%\tablecolumns{<num>}
\tablecaption{AGN candidates from IRAC colors\label{iracagntab}}
\tablehead{\colhead{MIPS ID} & \colhead{RA (J2000)} & \colhead{DEC (J2000)} & \colhead{log(I4/I2)} &\colhead{log(I3/I1)} &\colhead{z}&\colhead{Notes}\\ \colhead{} & \colhead{(deg)} & \colhead{(deg)} & \colhead{} &\colhead{} &\colhead{}&\colhead{}}
\startdata
4027 & 203.787811 & 40.988712 & 0.2142 & 0.6035 & 0.2369&\\
4481 & 203.981293 & 41.024662 & -0.1219 & 0.1628 & 0.2351&\\
5035 & 204.152267 & 41.076385 & -0.1493 & 0.01325 & 0.2345&\\
5107 & 204.037750 & 41.085625 & -0.1472 & 0.143 & 0.2374&\\
5188 & 203.938751 & 41.089951 & -0.1200 & 0.0128 & 0.2378&\\
4188 & 203.833618 & 41.001053 & -0.6362 & -0.4442 & 0.2280 & C1\\
5035 & 204.152267 & 41.076385 & -0.1211 & -0.1493 & 0.2345 & F1\\
\enddata
\tablenotetext{1.}{Errors on the IRAC magnitudes are typically at the 10\% level.}
\end{deluxetable}

\subsubsection{Quasars at the outskirts of the superstructure}

At low redshifts, z$<$0.3, radio quiet quasars are known to trace the large scale structure of the universe \citep{soc04}. They are usually found at the outskirts of regions of high galaxy density, like rich clusters. We therefore cross-identify the quasar catalog of \citet{ver06} to search for such objects in and around the Abell~1763 supercluster.

\begin{figure*}
\epsscale{1.8}
\plotone{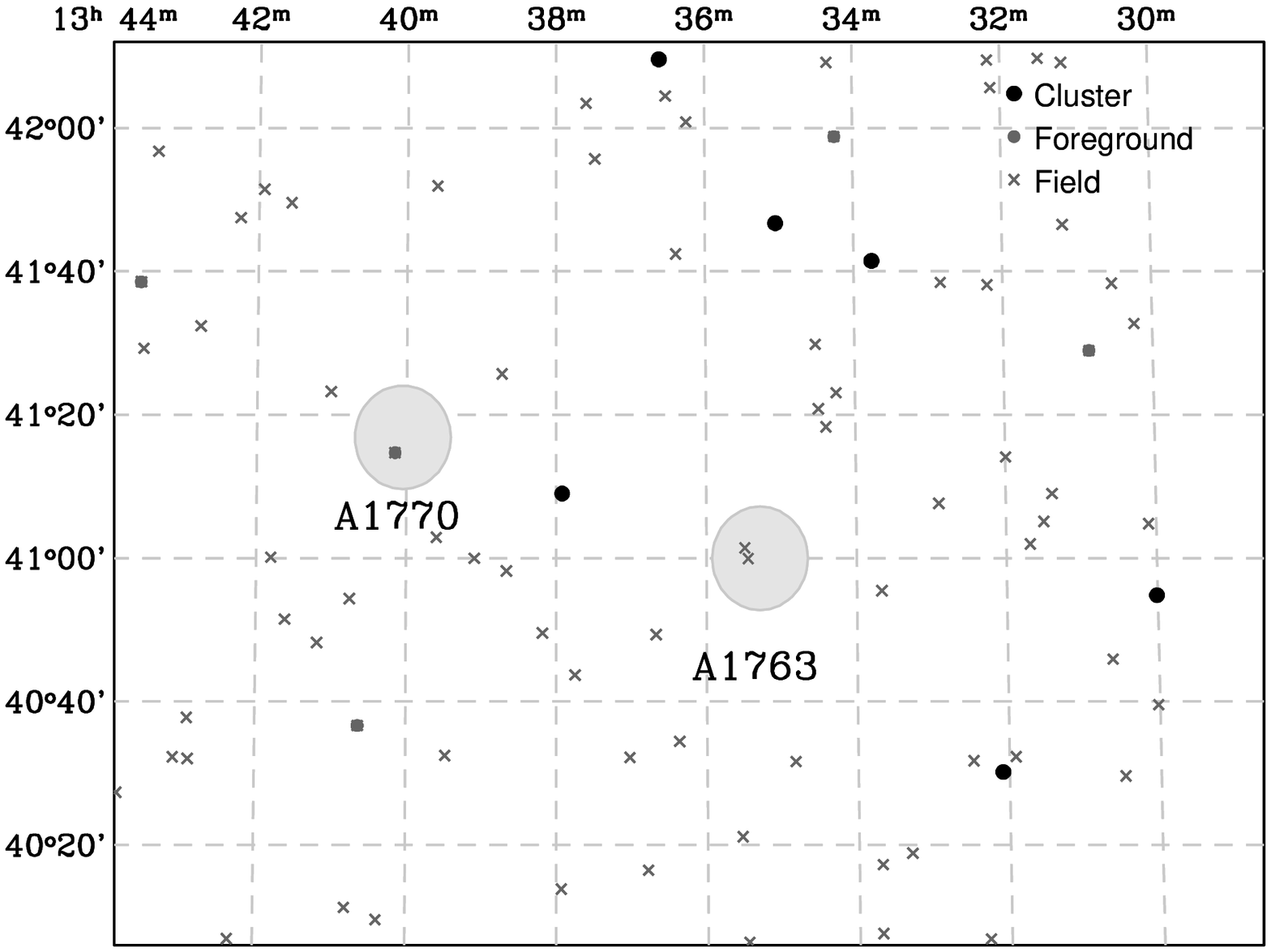}
\caption{{\bf Quasars near the Abell 1763 field.} The positions of quasars on the sky. The grey dots refer to those quasars with 0.15$<$z$<$0.2, and the black dots represent those with 0.2$<$z$<$0.25. Quasars outside of these redshift windows are shown with x's. The positions of Abell~1770 and Abell~1763 are shown as grey circles.}\label{quashist}\label{xyquas}
\end{figure*}

There are 85 quasars in the smaller region of 13h30m~$<$~$\alpha$~$<$~13h~44m, and 40d~40m~$<$~$\delta$~$<$~42d~20m, eleven of these have redshifts between 0.15 and 0.25. These quasars are listed in Table~\ref{quasartab} and shown in Figure~\ref{quashist}. The quasars which are spectroscopic cluster members are low luminosity quasars, having absolute V magnitudes that range between -20.2 and -22.7.

All quasars avoid the inner virial radius of the clusters with quasar ID 1322 found on the outer edge of Abell~1770, and ID 1259 in the middle of the galaxy filament. The latter is the ROSAT point source F3, positioned along the filament in Figure~\ref{radcov}.

\begin{deluxetable}{lccccc}
\tabletypesize{\scriptsize}
\tablewidth{0pt}
%\tablecolumns{<num>}
\tablecaption{Quasar candidates from \citet{ver06}\label{quasartab}}
\tablehead{\colhead{ID} & \colhead{RA (J2000)} & \colhead{DEC (J2000)} & \colhead{z} &\colhead{m$_{V}$}&\colhead{Note} \\\colhead{} & \colhead{(deg)} & \colhead{(deg)} & \colhead{} &\colhead{}&\colhead{}}
\startdata
1014 & 202.502917 & 40.913891 & 0.248 & 18.95 &\\
1031 & 202.713750 & 41.482224 & 0.182 & 17.63 &\\
1068 & 203.022923 & 40.502777 & 0.240 & 19.41 &\\
1119 & 203.439586 & 41.690834 & 0.225 & 18.14 &\\
1137 & 203.562082 & 41.980000 & 0.168 & 19.50 &\\
1173 & 203.762495 & 41.778332 & 0.225 & 19.96 &\\
1227 & 204.152922 & 42.159447 & 0.224 & 17.40 &\\
1259$^{a}$ & 204.480835 & 41.150002 & 0.231 & 18.76& F3  \\
1306$^{b}$ & 205.039171 & 41.244999 & 0.167 & 18.62& \\
1322 & 205.159171 & 40.610832 & 0.161 & 17.25 &\\
1408 & 205.894993 & 41.642502 & 0.164 & 18.56 &\\
\enddata
\tablenotetext{a}{This galaxy is labeled as F3 elsewhere in text.}
\tablenotetext{b}{Located on the outer side of A1770.}
\end{deluxetable}

\subsection{Starburst candidates and Radio SFRs}

The previous section has shown us that most of the super cluster members are SB but that a few AGN exist  In this section we use the optical emission lines to look into the AGN/SB nature of the members as well as quantify the amount of star formation and compare this to the amounts calculated from the FIR and radio emission.

\subsubsection{Candidates}

A map of the galaxies which are spectroscopically confirmed cluster members for which we can measure q$_{24}$ is shown in Figure~\ref{radcov}. For the few cluster members with high radio flux densities, the SB still dominate, but he fraction of AGN present is much higher. All radio bright SB and AGN are all projected along or towards the filament. The SB are all further from the center of Abell~1763, and 2 of 3 of the AGN are inside the cluster core. 

\subsubsection{Optical emission lines}

\begin{deluxetable}{lccccc}
\tabletypesize{\scriptsize}
\tablewidth{0pt}
%\tablecolumns{<num>}
\tablecaption{Cluster member optical emission line ratios\label{optlineagntab}}
\tablehead{\colhead{RA (J2000)} & \colhead{DEC (J2000)} & \colhead{log(OIII/H$\beta$)} &\colhead{log(NII/H$\alpha$)} &\colhead{z}&\colhead{Notes}\\ \colhead{(deg)} & \colhead{(deg)} & \colhead{} &\colhead{} &\colhead{} &\colhead{}}
\startdata
203.925201 & 41.039402 & 1.0253  &  0.0033 & 0.2181 &   C2\\ 
204.152267 & 41.076385 &$>$0.7731 & -0.1130 & 0.2345 &   F1\\
204.179306 & 41.101704 & -0.3310 & -0.3774 & 0.2316 &   F2\\ 
204.480927 & 41.150002 & -0.2867 & -0.3731 & 0.2311 &   F3, broad lines\\ 
204.452316 & 41.166481 & -0.5798 & -0.2580 & 0.2564 &   F4\\ 
204.265793 & 41.199951 & -0.5166 & -0.1793 & 0.2576 &   F5\\ 
204.194122 & 41.219238 &  0.4197 & -0.0764  & 0.2330 &   F6\\ 
\enddata
\end{deluxetable}

\begin{figure*}
   \center
  \epsscale{2.0}
     \plotone{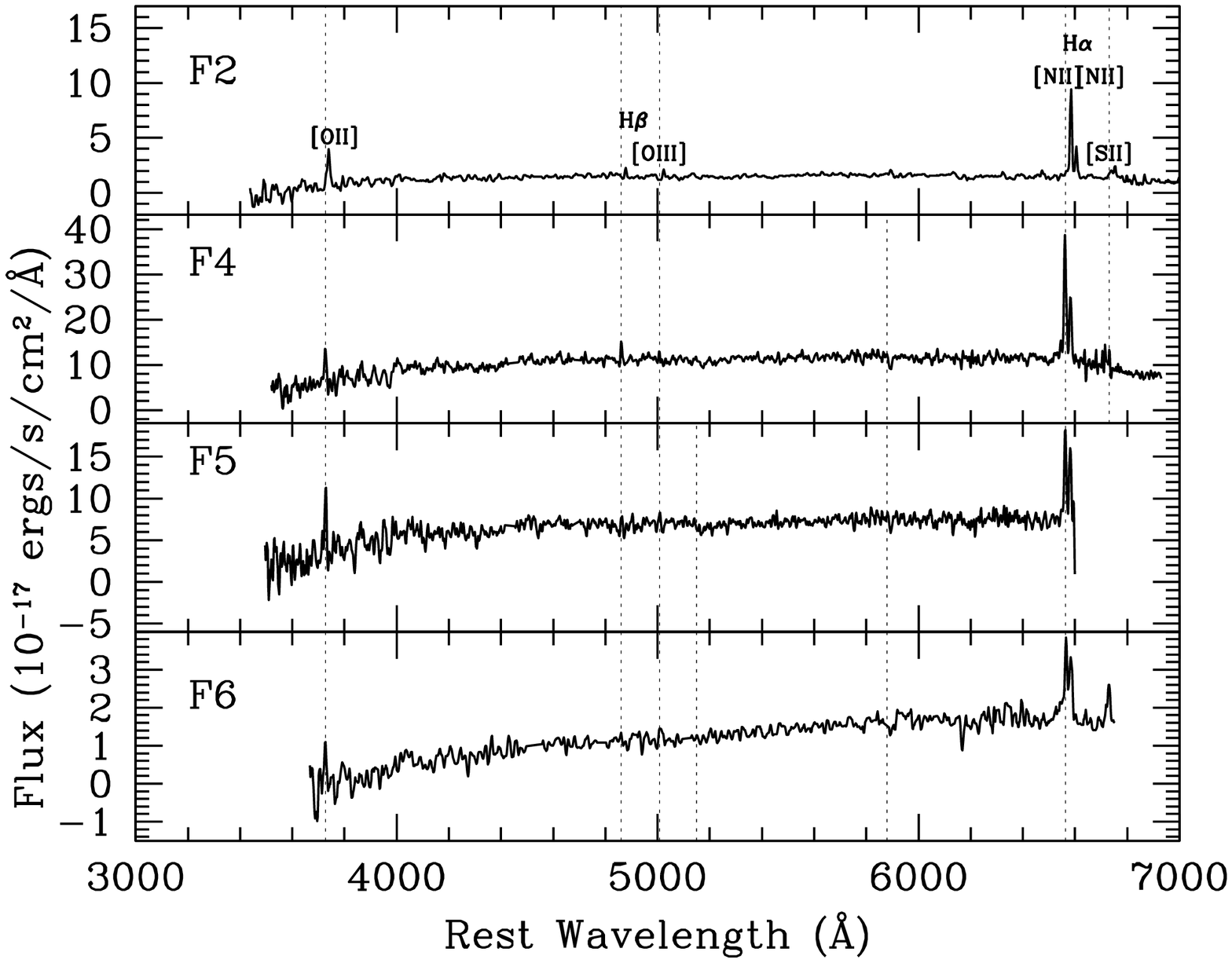}
    \caption[]{{\bf SB Galaxy spectra.}  F2, F4, F5, and F6 (as on Figure~\ref{radcov}) all have strong H$\alpha$ emission.
\label{filsbspec}}
\end{figure*}

\begin{figure*}
   \center
  \epsscale{2.0}
     \plotone{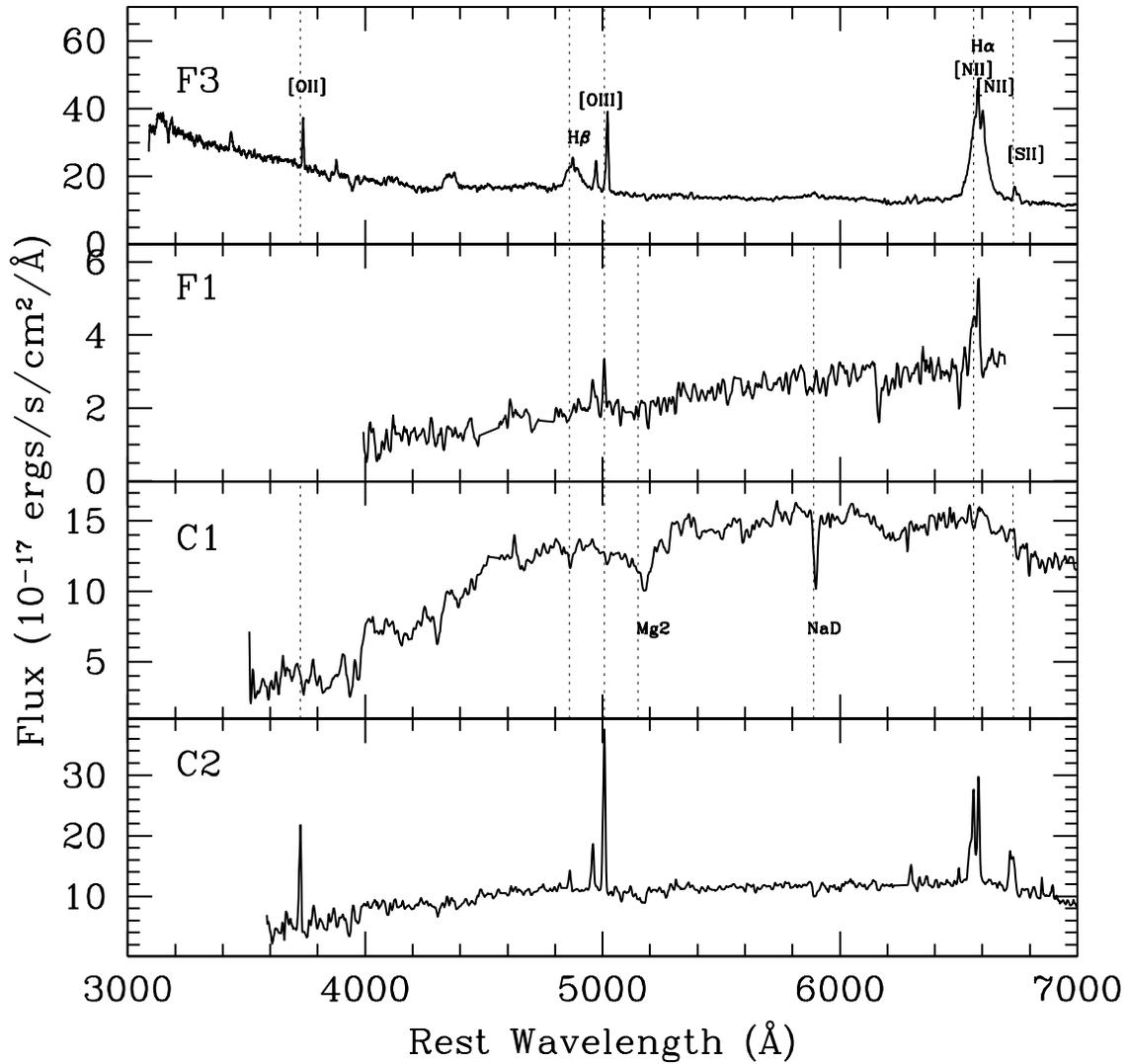}
    \caption{{\bf Bright AGN in the Supercluster.} AGN Galaxy spectra. From top to bottom,  F3 (ROSAT point source and known quasar), F1 (double lobe radio source), C1 (WAT), and C2 (XMM point source),  as labeled on Figure~\ref{radcov}. F3 shows broad emission lines, and C2 shows high [NII] to H$\alpha$ ratios, typical of AGN.
\label{filspec}}
\end{figure*}

\begin{figure}
\epsscale{0.8}
\plotone{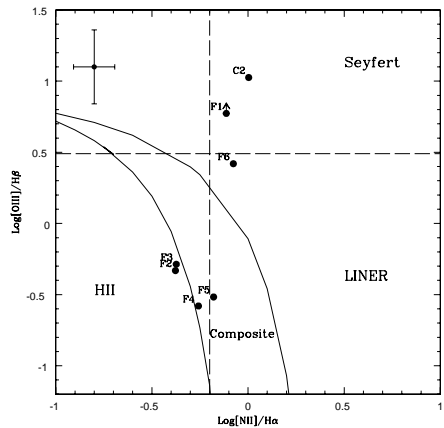}
\caption{{\bf BPT diagram.} HII regions, Seyferts, LINERs, and a possible composite of SB and AGN are separated on the BPT diagram. The emission line ratios are listed in Table~\ref{optlineagntab}. The sources plotted are those spectroscopically confirmed cluster members with measurable emission lines. Four lie in or near the composite region and two are Seyfert galaxies, and one is on the border of the LINER/Seyfert cut. F3 lies in the composite region but does have broad emission lines characteristic of AGN. The upper curve is the theoretical separation between SB and AGN from the population synthesis models of \citet{kew01}, and the bottom curve is from the observations of \citet{kau03}. The datapoint in the top left corner shows errorbars based on the average uncertainty in the emission line measurements.}\label{bptAGN}
\end{figure}

Figure~\ref{filsbspec} shows optical spectra of the four which have high H$\alpha$ to [NII] ratios, as expected for SB or composite galaxies. Figure~\ref{filspec} shows the three optical spectra indicative of AGN, as well as the optical spectra for the bright radio AGN central WAT.  F3 has fairly high H$\alpha$ to NII, however the lines are broad indicative of a type 1 AGN, it was also the ROSAT X-ray source.  These spectra have been processed with in house IDL scripts that correct for the bias, cosmic rays, flat-fielding, perform wavelength calibration, sky subtraction, atmospheric extinction, and correct for internal galactic extinction. The details of the data reduction process can be found in \citet{mar07}. We measured emission lines by minimizing the chi square fit of mutiple gaussians (for deblending emissions and any absorption) and a line for the continuum. The emission line ratios for the cluster members with optical spectra are listed in Table~\ref{optlineagntab} and plotted on the diagnostic diagram of \citet{bpt81} (hereafter BPT diagram; see Figure~\ref{bptAGN}).

\subsubsection{Star formation rates}

\begin{deluxetable}{lccccccccccc}
\tabletypesize{\scriptsize}
\tablewidth{0pt}
%\tablecolumns{<num>}
\tablecaption{SFRs From Members with Radio and IR Fluxes\label{sbprop}}
\tablehead{\colhead{Flux$_{Peak}$} &\colhead{Flux$_{24}$}&\colhead{Flux$_{70}$}&\colhead{Flux H$_{\alpha}$}& \colhead{z$_{sp}$}&\colhead{(u$^{\prime}$-r$^{\prime}$)}&\colhead{SFR$_{Rad}$}&\colhead{SFR$_{M24}$}&\colhead{SFR$_{M70}$}&\colhead{SFR$_{SED}$}&\colhead{SFR$_{H\alpha}$}&\colhead{Notes}}
\startdata
$\mu$Jy/bm  &   $\mu$Jy &   mJy & 10$^{-15}$cgs &  &  &  M$\odot$/yr & M$\odot$/yr & M$\odot$/yr & M$\odot$/yr & M$\odot$/yr  \\ 
\hline
%  922.7 $\pm$   283.1 &  1554.4 $\pm$    14.7 &   - & 11.66 $\pm$ 0.34  &0.2181 &   1.819 &      65 &      7.0& 2.7 & - & 3.21 &C2\\ 
  224.4 $\pm$    27.9 & 12110.8 $\pm$    65.8 & 56.5 $\pm$   0.8 & 3.06 $\pm$ 0.08     & 0.2316   &   1.798 &      14  &     94.9 & 28.9 & 49.8 & 3.7  & F2\\ 
  291.6 $\pm$    48.7 &  4262.4 $\pm$   112.1 &   -              & $<$140.1 & 0.2311  &   0.781 &      21 &     27.3   & - & 60.3 & $<$170.9     &F3\\
  419.4 $\pm$    44.3 &  1899.3 $\pm$    84.2 & 27.6 $\pm$   0.8 & 11.03 $\pm$ 0.26    & 0.2564 &   2.377 &      35 &     14.2    & 14.2 & 20.0 & 13.5      &F4\\ 
  277.7 $\pm$    49.9 &  1799.7 $\pm$    19.2 & 38.1 $\pm$   0.6 & 1.42  $\pm$ 0.21    & 0.2576 &   2.408 &     136 &     13.5    & 22.4 & 32.8 & 1.7     &F5\\ 
  219.5 $\pm$    61.4 &   816.3 $\pm$    17.5 & 25.5 $\pm$   0.8 & 2.25 $\pm$ 0.09    & 0.2330 &   2.535 &      51 &      3.9    & 10.7 & 15.4 & 2.7      &F6\\ 
\hline
84 & 700.2 $\pm$ 30.0 & - & - & 0.2319 & 2.264  & 6.4 & 3.2 & - & - & - &88 in A\\
369 & 831.7 $\pm$ 32.9 & - & - & 0.2325 & 2.222 & 30.5 & 4.0 & - & - & - &96 in B\\
\enddata
\end{deluxetable}
%OLD awk '($25>0&&($25/$5>1))||($27>0&&($27/$5>100)){printf "%d & %7.1f & %7.1f & %7.3f & %7.3f & %7.3f & %7.3f & %7.3f & %7.3f & %6.4f & %7.2f & %7.0f & %8.1f \\\\ \n", $1,$5/1000,$6/1000,$11,$15,$25/1000,$26/1000,$27/1000,$28/1000,$24,$32,$33,$34}' A1763_RadProp.ASC
%awk '$21<=3.0&&$24>=0.21&&$24<=0.26&&(($25>0&&($25/$5>0.28))||($27>0&&($27/$5>46.8))){printf "%7.1f \\pm %7.1f & %7.3f \\pm %7.3f & %7.3f \\pm %7.3f & %6.4f & %7.3f & %7.0f & %8.1f \\\\ \n", $5,$6,$25,$26,$27,$28,$24,$11-$15,$33,$34}' A1763_RadProp.ASC

SFRs based on radio observations are determined for members on the radio-FIR correlation. We include F2, F3 (with notes), F4, F5 and F6. We calculate the star formation rates based on radio luminosity \citep{yun01,bel03,gar09} and compare them to those based on the MIPS~24$\mu$m emission from \citet{rie09}, the MIPS~70$\mu$m from \citet{cal10}, and those based on a total infrared luminosity. Equation~(6) of \citet{gar09}, assuming a spectral index of $\alpha$=0.8, gives the SFRs listed in Table~\ref{sbprop}. To calculate the SFR from the MIPS~24$\mu$m luminosity, we use Equation~(14) of \citet{rie09}. We interpolate between bordering redshifts to obtain the proper coefficients A(z) and B(z) from their Table~1. Rates based on total FIR luminosities are calculated from best fit SEDs to the \citet{pol07} empirical templates and GRASIL population synthesis models \citep{sil98}. We then use the relationship from \citet{ken98} for the star formation rates. We use these same SED fits to deduce the rest MIPS~70$\mu$m flux from the data in order to be able to use the \citet{cal10} relationship. 

As these galaxies also have optical emission lines, we include the value of the extinction-corrected H$\alpha$ flux in Table~\ref{sbprop}. The H$\alpha$-derived SFRs included were derived using the formula of \citet{ken98}. The H$\alpha$ derived values tend to be lower than those calculated from the other methods. Whereas the radio and MIPS derived rates are based on observations of about the same 5$^{\prime\prime}$ beam, the H$\alpha$ fluxes are observed though the smaller 2$^{\prime\prime}$ diameter WYIN fibers. Because we do not know the the physical extent of the H$\alpha$ emission, we do not include an aperture correction to the H$\alpha$ fluxes. We note that if the line emission continues out to the same extent as the beam width of the radio, then a factor of 6.25 could be applied. 

The radio derived rates are usually below but similar to those derived from MIPS, however, F5 and F6 have high values of radio SFRs compared to those from MIPS and optical lines. This is not unlike the difference found between the radio and MIR (based on PAH luminosity) discrepancies found in \citet{sar09} who studied SFRs in infrared galaxies. Incidents where the radio SFR is significantly higher than the SFRs based on other indicators may be due to contaminating emission from a AGN. F6 has optical emission line ratios suggesting a possible Seyfert. Also, F3 has broad optical emission lines and is a ROSAT X-ray source so likely contaminated by an AGN as well. However, post starburst activity is commonly seen in these types of galaxies \citep{van06}.

  \begin{figure*}
   \center
  \epsscale{2.0}
     \plotone{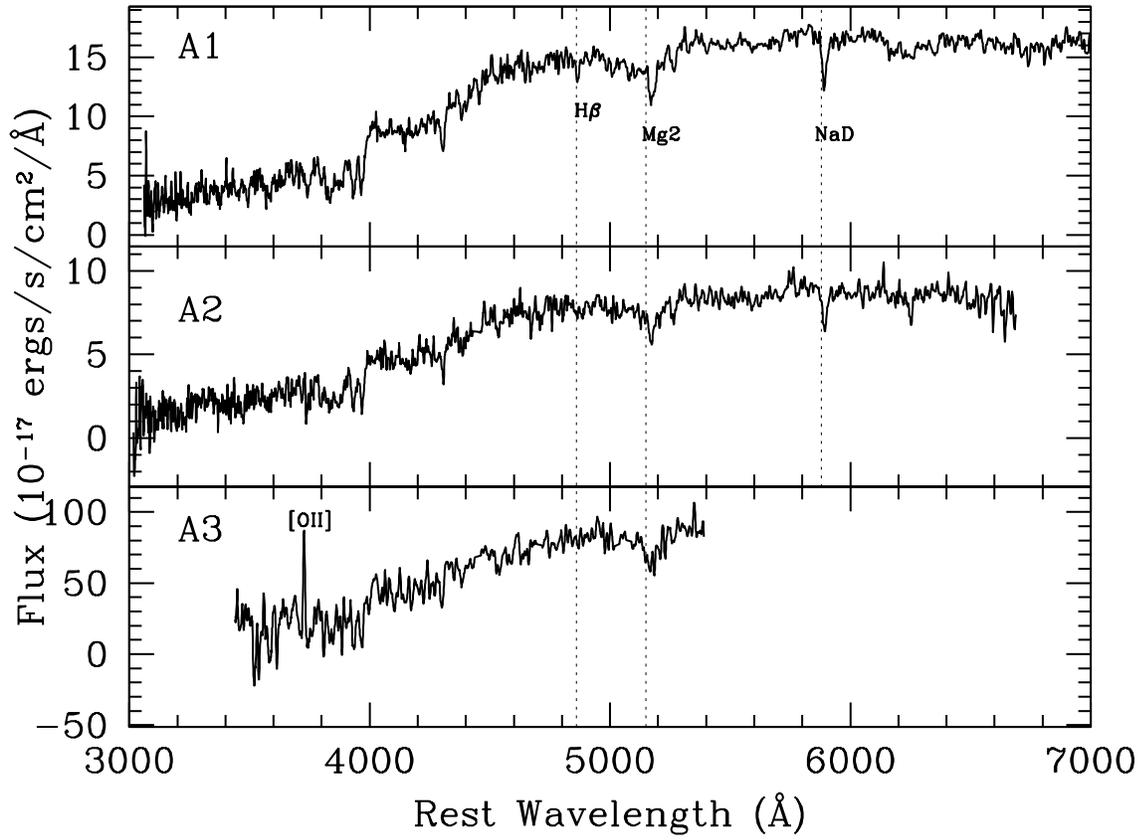}
    \caption[]{{\bf Absorption Line Galaxies.}  A1, A2, and A3 (as on Figure~\ref{radcov}) all have absorption lines typical of elliptical galaxies.
\label{filabsspec}}
\end{figure*}

We examined the optical spectra for the few spectroscopically confirmed supercluster galaxies detected in the radio without MIPS detections (Figure~\ref{filabsspec}) to compare the H$\alpha$ derived SFRs. These are the dark blue points of Figure~1 that are not labeled as either SF or AGN. However, for all cases there is no H$\alpha$ or H$\beta$ emission, but rather absorption. These galaxies may be regular ellipticals, or may reflect their dusty nature where IR and radio observations are important. We are currently preparing a more comprehensive analysis of all the supercluster spectra (Fadda, et al., in preparation).

\subsection{Bright radio-excess AGN in the field}

\begin{figure}
   \center
  \epsscale{1.0}
     \plotone{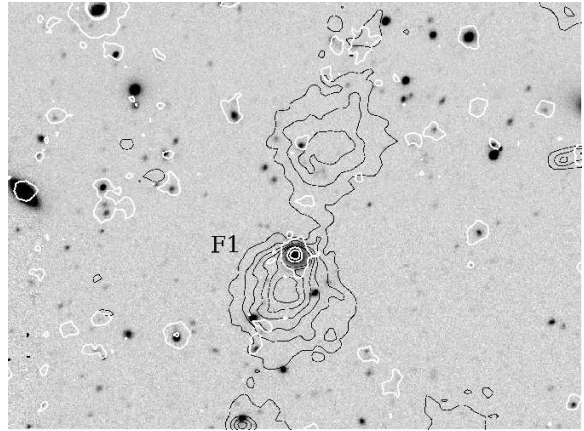}
    \caption[]{{\bf Bright Filament AGN}  The R-band image surrounding the radio-excess galaxy is shown in negative grey-scale. The galaxy is located at 13:36:36.2,41:04:47.3 (J2000) along the filament and NE of the BCG. The thin grey contours show the large symmetric lobes of radio emission and the thick white contours show the MIPS 24$\mu$m emission. This is labeled as F1 on Figure~\ref{radcov} and Figure~\ref{filspec}.
\label{bragn}}
\end{figure}

We observe two member AGN with radio flux $>$1.0$\,$mJy/bm. The first is the BCG at the center of Abell~1763 and the second is F1 which is located along the filament at 13:36:36.2 +41:04:47.3 (J2000), $\sim$15.3$\,$arcmin from the BCG. 

\subsubsection{The WAT BCG at the center of Abell~1763}

The BCG of Abell~1763 is a cD galaxy which is slightly offset from the X-ray peak of 18$\,$arcsec (63 kpc). This galaxy is well documented as a wide-angle tailed galaxy and radio images have already been published in \citet{owe92} and \citet{owe97} at the same resolution as our data, as well as in \citet{har04} at higher resolution and lower frequencies. The radio-FIR fluxes assure us that it is indeed a radio-excess AGN. We include the optical spectrum of this galaxy in Figure~\ref{filspec}, which shows a characteristic spectrum of a late type galaxy, common for cDs. The difference in the optical spectra illustrates the importance of using multiple techniques to identify AGN in cluster galaxies.

\subsubsubsection{Physical properties of the WAT}

Using the published spectral index of this WAT from \citet{owe97} of 1.08, we can calculate several physical properties of the WAT which enable us to further elucidate the properties of the surrounding ICM. First, we calculate the total luminosity of the system as in \citet{ode87} and assume the full radio spectrum to span from 10$\,$MHz to 100$\,$GHz as in \citet{smo07}. Taking the redshift to be 0.22 which leads to a luminosity distance of 1057$\,$Mpc, we derive L$_{tot}$=6.89$\times$10$^{43}$erg/s.

Magnetic fields, relativistic particles, and thermal pressure all contribute to the internal pressure in the radio jets \citep{pac70}, and this leads to a method for estimating the contribution of the magnetic fields \citep{pac70,ode87,smo07}. Assuming cylindrical symmetry and using the width of the minor axis of the clean beam for the height of the jet (16.6$\,$kpc), as well as a volume filling factor of 1, we assume that the energy of relativistic electrons equals that from relativistic protons and calculate the minimum magnetic field in the southern jet. Here, the flux density is 0.36$\,$Jy/Beam. The value for the magnetic field is 24.4$\,$$\mu$G, the minimum energy density is therefore 5.5$\times$10$^{-11}$dyn$\,$cm$^{-2}$, and the minimum pressure in the jet is therefore 1.8$\times$10$^{-11}$dyn$\,$cm$^{-2}$.

We can calculate the particle lifetime using the minimum magnetic field calculated above along with the magnetic field of the CMB radiation, which is 4(1+z)$^{2}$ \citep{ode87}. For an electron radiating at 1.4$\,$GHz, the lifetime is $\sim$6$\,$Myr. As the distance between the core and the hotspots in the jets is 20$\,$kpc \citep{har04}, a jet speed can be estimated of 0.01c. Jet speeds in other WATs have been measured to be 0.04-0.06c \citep{jet06,smo07}. Such high speeds are thought to keep the jet resistant to ram pressure as the BCG is near the bottom of the cluster potential well and should not be traveling at high speeds with respect to the ICM. 

\citet{har04}, \citet{jet06} and \citet{smo07} have found that relative speeds as low as 300-500km$\,$s$^{-1}$ are enough to bend WATs embedded in the ICM. From optical spectra of the BCG, and of hundreds of other galaxies in A1763, we have measured the velocity offset of the BCG from the cluster mean. In Paper~0, we published a projected offset of $\sim$260$\,$km$\,$s$^{-1}$. Considering the true velocity could be higher depending on the true direction of motion, this velocity is consistent with those calculated in the hydrodynamic models.

\subsubsection{A bent double lobe radio galaxy along the supercluster filament}

Figure~\ref{bragn} shows the R-band image of F1 with radio and 24$\,\mu$m contours overlaid. This source is well below both the 24$\,\mu$m and 70$\,\mu$m radio-FIR correlations. We note that this source is classified as an AGN whether or not the flux from the lobes is included.  Figure~\ref{filspec} shows the corresponding optical spectrum (F1, also discussed in Paper~0). High [NII] emission relative to H$\alpha$ is seen, a characteristic of AGN or LINER activity. The bright radio core of this double lobe source appears to be at the edge of the brighter lobe, and the fainter lobe is more extended, implying that we may be viewing a bent double lobe source on an angle. A full discussion of this source, its likely bent morphology, and its implications for the surrounding intra-filament medium  are detailed in a separate work (Edwards et al. 2010, in preparation).

\section{Discussion}

\subsection{The AGN and starburst fractions}

The majority of bright FIR cluster sources are SB systems. We have determined this using the radio-FIR correlation, the IRAC color diagnostic and through matching to X-ray point source and known quasar catalogs. However, for the few sources that have bright MIPS and bright Radio emission as well, the fraction of AGN sources rises highly to nearly 50\%. This is consistent with the results of \citet{smo08} who find roughly 50\% of sub-mJy radio sources in the COSMOS field at redshifts between 0.2-0.3 are AGN and roughly half are SB.

% awk '$25>=0&&($25/$5)>=1{print $0}' A1763_Radio_SM.ASC | wc -l
% awk '(sqrt(((203.9195-$2)*(203.9195-$2))*0.75+((41.06278-$3)*(41.06278-$3)))<0.36)&&$4>4000{print $0}' fulltabRelzAUG08.ASC | wc -l
% awk '$25>=0&&($25/$5)>=1&&($24<=0.26&&$24>=0.21){print $0}' A1763_Radio_SM.ASC | wc -l
%awk '$25>=0&&($25/$5)<=1{print $24,$22}' A1763_Radio_SM.ASC

\subsubsection{Radio Galaxies with MIPS 24$\mu$m $<$ 4$\,$mJy are SB}\label{stacksec}

We already found at the average of our stacked low-radio flux MIPS 24$\mu$m selected sample was described by starburst activity. But are we missing a large population of AGN with FIR emission below our MIPS 24$\mu$m sensitivity? To estimate the number of AGN we may be missing, we study the radio luminosity function, particularly that of cluster galaxies. \citet{yun01} construct a radio luminosity function of infrared selected galaxies out to a redshift of 1.5. They find that starbursts and AGN show a bimodal distribution, with AGN having 1.4$\,$GHz luminosities over 10$^{23}$ W$\,$Hz$^{-1}$ and starbursts having lower radio luminosities. By studying over 2000 SDSS galaxies, \citet{bes05} also find that the starburst galaxies dominate over the AGN population below 1.4$\,$GHz luminosities of 10$^{23}$ W$\,$Hz$^{-1}$. However, the cluster AGN number density is much higher than in the field \citep{lin07}, so it would be prudent to look at radio luminosity functions built on cluster galaxies. \citet{mil02} constructed a radio luminosity function based on galaxies in nearby galaxy clusters, also finding that star formation dominates in galaxies with radio luminosities less than 5$\times$10$^{22}$W$\,$Hz$^{-1}$. According to this, the galaxies below our detection limits (10$^{22}$W$\,$Hz$^{-1}$) should be star forming.

However, recent work by \citet{mil09} on the Coma cluster has found some surprising results. Though the authors give the caveat of a possible selection bias, they find fewer numbers of star forming galaxies than AGN at very low radio luminosities. By constructing a very deep radio luminosity function, the authors indeed find that low luminosity radio sources ($\sim$log~L$_{1.4GHz}$=22.2) are dominated by star forming systems, but at the very low radio luminosities ($\sim$log~L$_{1.4GHz}$$<$21), the AGN contribution to the radio luminosity function again becomes important. Such low luminosities are below our radio detection limits, so if A1763 has a similar radio luminosity function to Coma, we may in fact be missing some low-luminosity AGN. For sources at the redshift of A1763, log~L$_{1.4GHz}$$\sim$21, corresponds to a radio flux of 9.6$\,$$\mu$Jy - much below our radio detection limit. Following the FIR-radio correlation, and assuming it is valid at these lower luminosities, the corresponding MIPS 24$\mu$m flux for an AGN would have to be 96$\,$$\mu$Jy or below. Such a value is just below the 3$\sigma$ depth of 120$\,$$\mu$Jy for our MIPS observations, so it is possible our observations could include some of these galaxies, but they should not form a large population of the MIPS-selected cluster galaxies. In fact, there are 133 (out of 10876) MIPS sources with 24$\mu$m flux less than 100$\,\mu$Jy, but only two are spectroscopically confirmed cluster members. 

\subsection{Activity as a function of environment}

Table~\ref{ngalenv} lists the number and number density of active galaxies in the filament with radio detections (which includes the area of both clusters and a cylinder the width of approximately the virial radius of Abell 1763, grey circles from Figure~\ref{radcov}). The number density of radio and radio+MIPS detected galaxies inside and outside of the filament is approximately the same, 165 and 169 deg$^{-2}$, respectively. However, when looking at only the superstructure members, there is a higher number density of active galaxies inside the filament than in the outskirts and field (MIPS~SF and MIPS~AGN in Table~\ref{ngalenv}). In fact, this region is dominated by the starburst candidates and not the radio-excess AGN as the number density for star forming galaxies in the filament is higher than in the outskirts, whereas the number of AGN is higher in the outskirts. This shows that the filament is more active than the local field and outskirts and that it is not the radio-excess AGN that dominate the activity. It is remarkable that 9 of 11 of the quasars at the cluster redshift are found on the outskirts of the superstructure. 

\begin{deluxetable}{lcccccc}
\tabletypesize{\scriptsize}
\tablewidth{0pt}
%\tablecolumns{<num>}
\tablecaption{Number of active galaxies\label{ngalenv}}
\tablehead{\colhead{Num of Radio Galaxies} & \colhead{Filament} & \colhead{Area(deg$^{2}$)} &\colhead{density (deg$^{-2}$)} &\colhead{Outside} &\colhead{Area(deg$^{2}$)}&\colhead{Density (deg$^{-2}$)}}
\startdata
	&230	   &0.36&	639 &	361&	 0.54	&	669\\
%+
%z$_{ph}$	&25	   &0.36&	69&	12   &    0.54   &       22\\
+
z$_{sp}$	&11	   &0.36&	33&	1    &    0.54*&		2\\
+
MIPS	&33	   &0.20&	165&	44   &    0.26	&	169\\
+
MIPS+
z$_{sp}$	&8	   &0.20&	45&	0    &    0.26*&		0\\
+
MIPS SF	&18	  & 0.20&	90&	4    &    0.26	&	7\\
+
MIPS SF+
z$_{sp}$	&5	 &  0.20&	35&	0    &    0.26*&		0\\
+
MIPS AGN	&22	&   0.20&	110&	37   &    0.26	&	69\\
+
MIPS AGN+
z$_{sp}$ 	&3&	   0.20&	15&	0    &    0.26*&		0\\

\enddata
\end{deluxetable}

%awk '$3<205.13292&&$3>203.70167&&($4<(0.236711*($3-203.81267)+41.131444)&&$4>(0.237611*($3-203.81276)+40.876444))&&($24>0.21&&$24<0.26)&&(($25>=0&&$25/$5>1)||($27>0&&$27/$5>100)){print $0}' A1763_Radio_SM.ASC | wc -l
%awk '$3<205.13292&&$3>203.70167&&($4>(0.236711*($3-203.81267)+41.131444)||$4<(0.237611*($3-203.81276)+40.876444))&&($25>=0||$27>=0){print $0}' A1763_Radio_SM.ASC

\subsection{Cluster scale morphology and the double lobe radio sources}\label{watsec}

The current study provides three additional pieces of evidence which add to our claims of an ongoing merger or large scale flow of cluster feeding galaxies made in Paper~0.

\begin{figure}[t]
  \epsscale{1.0}
     \plotone{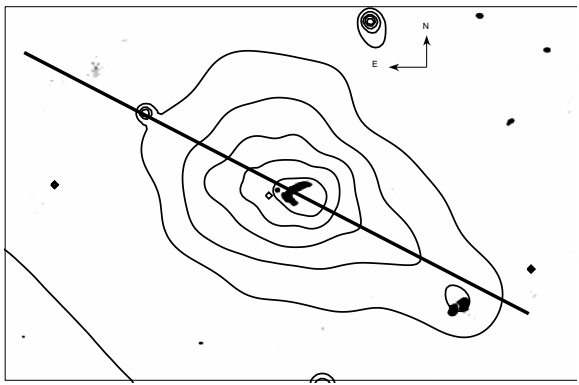}
    \caption{{\bf The BCG of A1763.} A $\sim$16 arcmin $\times$10 arcmin field centered on the BCG of A1763. This is a cD galaxy in the optical and appears as a bright WAT in the 1.4GHz images. The background image is the 1.4GHz data. This schematic shows X-ray contours from XMM overlaid. The X-ray emission is elongated in the same direction as the cluster feeding filament, and as the WAT tails. The black line shows the direction towards the cluster feeding filament galaxies and is angled at 25 degrees N of E.
\label{radx}}
\end{figure}

First, the XMM image reveals that the ICM is elongated along a southwest-northeast direction (Figure~\ref{radx}), with a cooler long tail of gas towards the southwest, similar to the case of Abell~85 \citep{dur03}. There is also a small displacement between the projected position of the BCG and the peak of  X-ray emission. This suggests that some kind of merging or flow has taken place along this direction.

Secondly, the Abell~1763 WAT consists of two tails, angled {\it away} from the direction of the cluster filament. This can be seen in Figure~\ref{radx} where we overplot the X-ray contours on our 1.4$\,$GHz image. Notice that the WAT is offset from the X-ray brightness peak by $\sim$100kpc (first noted in Paper~0). The two jets are oriented perpendicular to the large-scale morphology of the cluster X-ray emission, which itself has a projected elongation in the same direction as the cluster filament. The two large radio plumes have similar brightnesses, a ratio of 1.15 \citep{har04}, though the brightness ratio of the jets has been reported by \citet{jet06} to be much higher at 4.7. The plumes bend with the X-ray morphology and are parallel with the line of the cluster feeding filament galaxies. This data is consistent with the suggestion in \citet{dou08}, that the WAT morphology is an example of cool gas from the structure of the filament raining onto the cluster potential, at the location of the BCG. We take this as further evidence for the existence of the cluster feeding filament. In a future article we will include a discussion of the velocity dispersions and cluster components found from the analysis of the cluster galaxies optical spectra. Bent double lobe radio sources are rare, and that we have a second one in the same large scale structure may alluding to the dynamics in the the system. 

Finally, a score of lower luminosity quasars are found at the outskirts of the high galaxy density regions of the supercluster, with no quasars inside the cluster cores. This type of behavior in low luminosity quasars has previously been associated with merging clusters \citep{soc02,soc04}, at least for redshifts below z$\sim$0.3.

\section{Summary and conclusions}

We have examined 1.4$\,$GHz radio sources in the supercluster of Abell~1763-Abell~1770. We investigate the radio-FIR correlation, the location of X-ray point sources, the IRAC colors, and the location of known quasars to look at the frequency of occurrence of AGN in the sample of MIPS and radio selected galaxies. Taking advantage of radio upper limits derived from our data and MIPS fluxes from Paper~I, we conclude that the MIPS cluster galaxies that lie in the filament are mostly SB and not AGN.  We measure q$_{24}$ for 8 radio bright member galaxies. Five are classified SB galaxies, all along the filament. One AGN is located along the filament, and the other two AGN are inside the core of Abell~1763. For the galaxies classified as radio SB, we calculate the radio star formation rates and compare to those from the FIR and H$\alpha$ measurements. More redshifts of the radio galaxies would help increase our ability to discriminate between SB versus AGN inside and outside the superstructure. 

We calculate physical properties of the central WAT and of F1 and relate them to the large scale structure of this system mapped by XMM.  The BCG velocity is consistent with those induced in cluster-sub-cluster mergers and the morphology of the WAT and its orientation away from the cluster feeding filament suggests a relationship with the large scale flow of the cluster feeding filament and its ICM \citep{lok95,gom97}.  F1 is the first bent DLRS to be identified within a known supercluster filament. We suggest clusters which host a central WAT and or a source like F1 near to galaxy clusters may be good tracers for superclusters scale filaments.

There exist a number of quasars with redshifts consistent with those of the superstructure, at those on the outskirts of the system, a phenomenon previously associated with merging systems. Another method of selecting galaxy clusters with large filaments such as these would be to focus on superclusters near quasar triplets, which are rare. Deeper X-ray observations would help elucidate the large scale cluster dynamics.

If Coma, our closest rich cluster, is similar to other systems further out, it would be prudent for us to extend this study to even lower radio luminosities in order to fully uncover the contaminating AGN population to our study of SB in the A1763 filament. Although our {\em Spitzer} observations are likely not deep enough to find significant numbers of this radio-faint AGN population. The improved sensitivity of the EVLA and Herschel will make these types of studies possible.

\acknowledgments

We thank the scientific staff at the NRAO, in particular G. Van Moorsel for help with observation planning and F. Owen for help with data reduction techniques. We thank A. Biviano for his reading and comments on this manuscript and for constructing the FIR SEDs. We also thank the anonymous referee for extremely helpful comments that contributed to a much improved version of this paper.

 Support for this work was provided by NASA through an award issued by JPL/Caltech. This work is based in part on original observations using the Very Large Array operated by the NRAO. The National Radio Astronomy Observatory is a facility of the National Science Foundation operated under cooperative agreement by Associated Universities, Inc. This work is also based in part on archival observations obtained with XMM-Newton, an ESA science mission with instruments and contributions directly funded by ESA Member States and NASA and observations made with {\em Spitzer}, a space telescope operated by the Jet Propulsion Laboratory, California Institute of Technology, under a contract with NASA. Funding for the SDSS and SDSS-II has been provided by the Alfred P. Sloan Foundation, then Participating Institutions, the National Science Foundation, the US Department of Energy, NASA, the Japanese Monbukagakusho, the Max Planck Society, and the Higher Education Funding Council of England. The SDSS is managed by the Astrophysical Research Consortium for the Participating Institutions (see list at http://www.sdss.org/collaboration/credits.html). We have made use of the ROSAT Data Archive of the Max-Planck-Institut f\"ur extraterrestrische Physik (MPE) at Garching, Germany as well as the XMM-Newton Data Archive.

{\it Facilities:} Spitzer (IRAC), Spitzer (MIPS), VLA, XMM-Newton, WIYN (Hydra)

\bibliography{a1763}

\begin{thebibliography}{73}
\expandafter\ifx\csname natexlab\endcsname\relax\def\natexlab#1{#1}\fi

\bibitem[{{Appleton} {et~al.}(2004){Appleton}, {Fadda}, \& {Marleau}}]{app04}
{Appleton}, P.~N., {Fadda}, D.~T., \& {Marleau}, F.~R. e.~a. 2004, ApJS, 154,
  147

\bibitem[{{Baldwin} {et~al.}(1981){Baldwin}, {Phillips}, \&
  {Terlevich}}]{bpt81}
{Baldwin}, J.~A., {Phillips}, M.~M., \& {Terlevich}, R. 1981, PASP, 93, 5

\bibitem[{{Bell}(2003)}]{bel03}
{Bell}, E.~F. 2003, \apj, 586, 794

\bibitem[{{Best} {et~al.}(2005){Best}, {Kauffmann}, {Heckman}, {Brinchmann},
  {Charlot}, {Ivezi{\'c}}, \& {White}}]{bes05}
{Best}, P.~N., {Kauffmann}, G., {Heckman}, T.~M., {Brinchmann}, J., {Charlot},
  S., {Ivezi{\'c}}, {\v Z}., \& {White}, S.~D.~M. 2005, \mnras, 362, 25

\bibitem[{{Bond} {et~al.}(1996){Bond}, {Kofman}, \& {Pogosyan}}]{bon96}
{Bond}, J.~R., {Kofman}, L., \& {Pogosyan}, D. 1996, Nature, 380, 603

\bibitem[{{Burns} {et~al.}(1981){Burns}, {Gregory}, \& {Holman}}]{bur81}
{Burns}, J.~O., {Gregory}, S.~A., \& {Holman}, G.~D. 1981, ApJ, 250, 450

\bibitem[{{Calzetti} {et~al.}(2010){Calzetti}, {Wu}, \& {Hong}}]{cal10}
{Calzetti}, D., {Wu}, S., \& {Hong}, S., e.~a. 2010, ApJ, 714, 1256

\bibitem[{{Carter} \& {Read}(2007)}]{Carter07}
{Carter}, J.~A. \& {Read}, A.~M. 2007, \aap, 464, 1155

\bibitem[{{Ciliegi} {et~al.}(2003){Ciliegi}, {Zamorani}, {Hasinger}, {Lehmann},
  {Szokoly}, \& {Wilson}}]{cil03}
{Ciliegi}, P., {Zamorani}, G., {Hasinger}, G., {Lehmann}, I., {Szokoly}, G., \&
  {Wilson}, G. 2003, A\&A, 398, 901

\bibitem[{{Clowe} {et~al.}(2006){Clowe}, {Brada{\v c}}, {Gonzalez},
  {Markevitch}, {Randall}, {Jones}, \& {Zaritsky}}]{clo06}
{Clowe}, D., {Brada{\v c}}, M., {Gonzalez}, A.~H., {Markevitch}, M., {Randall},
  S.~W., {Jones}, C., \& {Zaritsky}, D. 2006, ApJL, 648, L109

\bibitem[{{Condon}(1992)}]{con92}
{Condon}, J.~J. 1992, ARA\&A, 30, 575

\bibitem[{{Condon}(1997)}]{con97}
---. 1997, \pasp, 109, 166

\bibitem[{{Coziol} {et~al.}(2009){Coziol}, {Andernach}, {Caretta},
  {Alamo-Mart{\'{\i}}nez}, \& {Tago}}]{coz09}
{Coziol}, R., {Andernach}, H., {Caretta}, C.~A., {Alamo-Mart{\'{\i}}nez},
  K.~A., \& {Tago}, E. 2009, AJ, 137, 4795

\bibitem[{{De Lucia} \& {Blaizot}(2007)}]{del07}
{De Lucia}, G. \& {Blaizot}, J. 2007, MNRAS, 375, 2

\bibitem[{{Donley} {et~al.}(2008){Donley}, {Rieke}, {P{\'e}rez-Gonz{\'a}lez},
  \& {Barro}}]{don08}
{Donley}, J.~L., {Rieke}, G.~H., {P{\'e}rez-Gonz{\'a}lez}, P.~G., \& {Barro},
  G. 2008, \apj, 687, 111

\bibitem[{{Douglass} {et~al.}(2008){Douglass}, {Blanton}, {Clarke}, {Sarazin},
  \& {Wise}}]{dou08}
{Douglass}, E.~M., {Blanton}, E.~L., {Clarke}, T.~E., {Sarazin}, C.~L., \&
  {Wise}, M. 2008, ApJ, 673, 763

\bibitem[{{Durret} {et~al.}(2003){Durret}, {Lima Neto}, {Forman}, \&
  {Churazov}}]{dur03}
{Durret}, F., {Lima Neto}, G.~B., {Forman}, W., \& {Churazov}, E. 2003, \aap,
  403, L29

\bibitem[{{Edwards} {et~al.}(2010){Edwards}, {Fadda}, {Biviano}, \&
  {Marleau}}]{edw10}
{Edwards}, L.~O.~V., {Fadda}, D., {Biviano}, A., \& {Marleau}, F.~R. 2010, AJ,
  139, 434

\bibitem[{{Eisenhardt} {et~al.}(2004){Eisenhardt}, {Stern}, \&
  {Brodwin}}]{eis04}
{Eisenhardt}, P.~R., {Stern}, D., \& {Brodwin}, e.~a. 2004, \apjs, 154, 48

\bibitem[{{Fadda} {et~al.}(2008){Fadda}, {Biviano}, {Marleau},
  {Storrie-Lombardi}, \& {Durret}}]{fad08}
{Fadda}, D., {Biviano}, A., {Marleau}, F.~R., {Storrie-Lombardi}, L.~J., \&
  {Durret}, F. 2008, \apjl, 672, L9

\bibitem[{{Finn} {et~al.}(2008){Finn}, {Balogh}, {Zaritsky}, {Miller}, \&
  {Nichol}}]{fin08}
{Finn}, R.~A., {Balogh}, M.~L., {Zaritsky}, D., {Miller}, C.~J., \& {Nichol},
  R.~C. 2008, \apj, 679, 279

\bibitem[{{Garn} {et~al.}(2009){Garn}, {Green}, {Riley}, \&
  {Alexander}}]{gar09}
{Garn}, T., {Green}, D.~A., {Riley}, J.~M., \& {Alexander}, P. 2009, \mnras,
  397, 1101

\bibitem[{{Gomez} {et~al.}(1997){Gomez}, {Ledlow}, {Burns}, {Pinkey}, \&
  {Hill}}]{gom97}
{Gomez}, P.~L., {Ledlow}, M.~J., {Burns}, J.~O., {Pinkey}, J., \& {Hill}, J.~M.
  1997, AJ, 114, 1711

\bibitem[{{Gonzalez} \& {Padilla}(2009)}]{gon09}
{Gonzalez}, R.~E. \& {Padilla}, N.~E. 2009, ArXiv 0912.0006

\bibitem[{{Griffith} \& {Stern}(2010)}]{gri10}
{Griffith}, R.~L. \& {Stern}, D. 2010, \aj, 140, 533

\bibitem[{{Hardcastle} \& {Sakelliou}(2004)}]{har04}
{Hardcastle}, M.~J. \& {Sakelliou}, I. 2004, MNRAS, 349, 560

\bibitem[{{Helou} {et~al.}(1985){Helou}, {Soifer}, \& {Rowan-Robinson}}]{hel85}
{Helou}, G., {Soifer}, B.~T., \& {Rowan-Robinson}, M. 1985, \apjl, 298, L7

\bibitem[{{Hickox} {et~al.}(2009){Hickox}, {Jones}, {Forman}, \&
  {Murray}}]{hic09}
{Hickox}, R.~C., {Jones}, C., {Forman}, W.~R., \& {Murray}, e.~a. 2009, \apj,
  696, 891

\bibitem[{{Imanishi} {et~al.}(2007){Imanishi}, {Dudley}, {Maiolino}, {Maloney},
  {Nakagawa}, \& {Risaliti}}]{ima07}
{Imanishi}, M., {Dudley}, C.~C., {Maiolino}, R., {Maloney}, P.~R., {Nakagawa},
  T., \& {Risaliti}, G. 2007, ApJS, 171, 72

\bibitem[{{Jetha} {et~al.}(2006){Jetha}, {Hardcastle}, \& {Sakelliou}}]{jet06}
{Jetha}, N.~N., {Hardcastle}, M.~J., \& {Sakelliou}, I. 2006, MNRAS, 368, 609

\bibitem[{{Jones} \& {Forman}(1984)}]{jon84}
{Jones}, C. \& {Forman}, W. 1984, \apj, 276, 38

\bibitem[{{Kauffmann} {et~al.}(1999){Kauffmann}, {Colberg}, {Diaferio}, \&
  {White}}]{kau99}
{Kauffmann}, G., {Colberg}, J.~M., {Diaferio}, A., \& {White}, S.~D.~M. 1999,
  MNRAS, 307, 529

\bibitem[{{Kauffmann} {et~al.}(2003){Kauffmann}, {Heckman}, {Tremonti},
  {Brinchmann}, {Charlot}, {White}, {Ridgway}, {Brinkmann}, {Fukugita}, {Hall},
  {Ivezi{\'c}}, {Richards}, \& {Schneider}}]{kau03}
{Kauffmann}, G., {Heckman}, T.~M., {Tremonti}, C., {Brinchmann}, J., {Charlot},
  S., {White}, S.~D.~M., {Ridgway}, S.~E., {Brinkmann}, J., {Fukugita}, M.,
  {Hall}, P.~B., {Ivezi{\'c}}, {\v Z}., {Richards}, G.~T., \& {Schneider},
  D.~P. 2003, MNRAS, 346, 1055

\bibitem[{{Kennicutt}(1998)}]{ken98}
{Kennicutt}, Jr., R.~C. 1998, ARA\&A, 36, 189

\bibitem[{{Kewley} \& {Dopita}(2002)}]{kew02}
{Kewley}, L.~J. \& {Dopita}, M.~A. 2002, \apjs, 142, 35

\bibitem[{{Kewley} {et~al.}(2001){Kewley}, {Dopita}, {Sutherland}, {Heisler},
  \& {Trevena}}]{kew01}
{Kewley}, L.~J., {Dopita}, M.~A., {Sutherland}, R.~S., {Heisler}, C.~A., \&
  {Trevena}, J. 2001, \apj, 556, 121

\bibitem[{{Lacy} {et~al.}(2004){Lacy}, {Storrie-Lombardi}, \& {Sajina}}]{lac04}
{Lacy}, M., {Storrie-Lombardi}, L.~J., \& {Sajina}, A., e.~a. 2004, \apjs, 154,
  166

\bibitem[{{Lagan{\'a}} {et~al.}(2008){Lagan{\'a}}, {Lima Neto},
  {Andrade-Santos}, \& {Cypriano}}]{Lagana08}
{Lagan{\'a}}, T.~F., {Lima Neto}, G.~B., {Andrade-Santos}, F., \& {Cypriano},
  E.~S. 2008, \aap, 485, 633

\bibitem[{{Lin} \& {Mohr}(2007)}]{lin07}
{Lin}, Y.-T. \& {Mohr}, J.~J. 2007, ApJS, 170, 71

\bibitem[{{Loken} {et~al.}(1995){Loken}, {Roettiger}, {Burns}, \&
  {Norman}}]{lok95}
{Loken}, C., {Roettiger}, K., {Burns}, J.~O., \& {Norman}, M. 1995, ApJ, 445,
  80

\bibitem[{{Marleau} {et~al.}(2007){Marleau}, {Fadda}, {Appleton},
  {Noriega-Crespo}, {Im}, \& {Clancy}}]{mar07}
{Marleau}, F.~R., {Fadda}, D., {Appleton}, P.~N., {Noriega-Crespo}, A., {Im},
  M., \& {Clancy}, D. 2007, ApJ, 663, 218

\bibitem[{{Miley} {et~al.}(1972){Miley}, {Perola}, {van der Kruit}, \& {van der
  Laan}}]{mil72}
{Miley}, G.~K., {Perola}, G.~C., {van der Kruit}, P.~C., \& {van der Laan}, H.
  1972, Nature, 237, 269

\bibitem[{{Miller} {et~al.}(2005){Miller}, {Nichol}, \& {Reichart}}]{mil05}
{Miller}, C.~J., {Nichol}, R.~C., \& {Reichart}, e.~a. 2005, AJ, 130, 968

\bibitem[{{Miller} {et~al.}(2009){Miller}, {Hornschemeier}, {Mobasher},
  {Bridges}, {Hudson}, {Marzke}, \& {Smith}}]{mil09}
{Miller}, N.~A., {Hornschemeier}, A.~E., {Mobasher}, B., {Bridges}, T.~J.,
  {Hudson}, M.~J., {Marzke}, R.~O., \& {Smith}, R.~J. 2009, AJ, 137, 4450

\bibitem[{{Miller} \& {Owen}(2002)}]{mil02}
{Miller}, N.~A. \& {Owen}, F.~N. 2002, AJ, 124, 2453

\bibitem[{{Navarro} {et~al.}(1995){Navarro}, {Frenk}, \& {White}}]{nav95}
{Navarro}, J.~F., {Frenk}, C.~S., \& {White}, S.~D.~M. 1995, MNRAS, 275, 56

\bibitem[{{O'Dea} \& {Owen}(1987)}]{ode87}
{O'Dea}, C.~P. \& {Owen}, F.~N. 1987, ApJ, 316, 95

\bibitem[{{Owen} \& {Ledlow}(1997)}]{owe97}
{Owen}, F.~N. \& {Ledlow}, M.~J. 1997, ApJS, 108, 41

\bibitem[{{Owen} {et~al.}(1995){Owen}, {Ledlow}, \& {Keel}}]{owe95}
{Owen}, F.~N., {Ledlow}, M.~J., \& {Keel}, W.~C. 1995, AJ, 109, 14

\bibitem[{{Owen} \& {Morrison}(2008)}]{owe08}
{Owen}, F.~N. \& {Morrison}, G.~E. 2008, AJ, 136, 1889

\bibitem[{{Owen} \& {Rudnick}(1976)}]{owe76}
{Owen}, F.~N. \& {Rudnick}, L. 1976, ApJL, 205, L1

\bibitem[{{Owen} {et~al.}(1992){Owen}, {White}, \& {Burns}}]{owe92}
{Owen}, F.~N., {White}, R.~A., \& {Burns}, J.~O. 1992, ApJS, 80, 501

\bibitem[{{Pacholczyk}(1970)}]{pac70}
{Pacholczyk}, A.~G. 1970, {Radio astrophysics. Nonthermal processes in galactic
  and extragalactic sources}

\bibitem[{{Peres} {et~al.}(1998){Peres}, {Fabian}, {Edge}, {Allen},
  {Johnstone}, \& {White}}]{per98}
{Peres}, C.~B., {Fabian}, A.~C., {Edge}, A.~C., {Allen}, S.~W., {Johnstone},
  R.~M., \& {White}, D.~A. 1998, MNRAS, 298, 416

\bibitem[{{Polletta} {et~al.}(2007){Polletta}, {Tajer}, {Maraschi},
  {Trinchieri}, {Lonsdale}, {Chiappetti}, {Andreon}, {Pierre}, {Le F{\`e}vre},
  {Zamorani}, {Maccagni}, {Garcet}, {Surdej}, {Franceschini}, {Alloin},
  {Shupe}, {Surace}, {Fang}, {Rowan-Robinson}, {Smith}, \& {Tresse}}]{pol07}
{Polletta}, M., {Tajer}, M., {Maraschi}, L., {Trinchieri}, G., {Lonsdale},
  C.~J., {Chiappetti}, L., {Andreon}, S., {Pierre}, M., {Le F{\`e}vre}, O.,
  {Zamorani}, G., {Maccagni}, D., {Garcet}, O., {Surdej}, J., {Franceschini},
  A., {Alloin}, D., {Shupe}, D.~L., {Surace}, J.~A., {Fang}, F.,
  {Rowan-Robinson}, M., {Smith}, H.~E., \& {Tresse}, L. 2007, \apj, 663, 81

\bibitem[{{Rieke} {et~al.}(2009){Rieke}, {Alonso-Herrero}, {Weiner},
  {P{\'e}rez-Gonz{\'a}lez}, {Blaylock}, {Donley}, \& {Marcillac}}]{rie09}
{Rieke}, G.~H., {Alonso-Herrero}, A., {Weiner}, B.~J.,
  {P{\'e}rez-Gonz{\'a}lez}, P.~G., {Blaylock}, M., {Donley}, J.~L., \&
  {Marcillac}, D. 2009, \apj, 692, 556

\bibitem[{{Sajina} {et~al.}(2005){Sajina}, {Lacy}, \& {Scott}}]{saj05}
{Sajina}, A., {Lacy}, M., \& {Scott}, D. 2005, \apj, 621, 256

\bibitem[{{Sakelliou} \& {Merrifield}(2000)}]{sak00}
{Sakelliou}, I. \& {Merrifield}, M.~R. 2000, MNRAS, 311, 649

\bibitem[{{Sargsyan} \& {Weedman}(2009)}]{sar09}
{Sargsyan}, L.~A. \& {Weedman}, D.~W. 2009, \apj, 701, 1398

\bibitem[{{Silk} \& {Rees}(1998)}]{sil98}
{Silk}, J. \& {Rees}, M.~J. 1998, A\&A, 331, L1

\bibitem[{{Smol{\v c}i{\'c}} {et~al.}(2007){Smol{\v c}i{\'c}}, {Schinnerer}, \&
  {Finoguenov}}]{smo07}
{Smol{\v c}i{\'c}}, V., {Schinnerer}, E., \& {Finoguenov}, A., e.~a. 2007,
  ApJS, 172, 295

\bibitem[{{Smol{\v c}i{\'c}} {et~al.}(2008){Smol{\v c}i{\'c}}, {Schinnerer}, \&
  {Scodeggio}}]{smo08}
{Smol{\v c}i{\'c}}, V., {Schinnerer}, E., \& {Scodeggio}, e.~a. 2008, ApJS,
  177, 14

\bibitem[{{S{\"o}chting} {et~al.}(2002){S{\"o}chting}, {Clowes}, \&
  {Campusano}}]{soc02}
{S{\"o}chting}, I.~K., {Clowes}, R.~G., \& {Campusano}, L.~E. 2002, \mnras,
  331, 569

\bibitem[{{S{\"o}chting} {et~al.}(2004){S{\"o}chting}, {Clowes}, \&
  {Campusano}}]{soc04}
---. 2004, \mnras, 347, 1241

\bibitem[{{Springel} {et~al.}(2005){Springel}, {White}, \& {Jenkins}}]{spr05}
{Springel}, V., {White}, S.~D.~M., \& {Jenkins}, A., e.~a. 2005, Nature, 435,
  629

\bibitem[{{Sutherland} \& {Saunders}(1992)}]{sut92}
{Sutherland}, W. \& {Saunders}, W. 1992, MNRAS, 259, 413

\bibitem[{{Umetsu} {et~al.}(2009){Umetsu}, {Medezinski}, {Broadhurst},
  {Zitrin}, {Okabe}, {Hsieh}, \& {Molnar}}]{ume09}
{Umetsu}, K., {Medezinski}, E., {Broadhurst}, T., {Zitrin}, A., {Okabe}, N.,
  {Hsieh}, B.-C., \& {Molnar}, S.~M. 2009, ArXiv e-prints

\bibitem[{{Vanden Berk} {et~al.}(2006){Vanden Berk}, {Shen}, {Yip},
  {Schneider}, {Connolly}, {Burton}, {Jester}, {Hall}, {Szalay}, \&
  {Brinkmann}}]{van06}
{Vanden Berk}, D.~E., {Shen}, J., {Yip}, C., {Schneider}, D.~P., {Connolly},
  A.~J., {Burton}, R.~E., {Jester}, S., {Hall}, P.~B., {Szalay}, A.~S., \&
  {Brinkmann}, J. 2006, AJ, 131, 84

\bibitem[{{V{\'e}ron-Cetty} \& {V{\'e}ron}(2006)}]{ver06}
{V{\'e}ron-Cetty}, M. \& {V{\'e}ron}, P. 2006, \aap, 455, 773

\bibitem[{{Vikhlinin} {et~al.}(2005){Vikhlinin}, {Markevitch}, {Murray},
  {Jones}, {Forman}, \& {Van Speybroeck}}]{vik05}
{Vikhlinin}, A., {Markevitch}, M., {Murray}, S.~S., {Jones}, C., {Forman}, W.,
  \& {Van Speybroeck}, L. 2005, ApJ, 628, 655

\bibitem[{{White}(2000)}]{whi00}
{White}, D.~A. 2000, MNRAS, 312, 663

\bibitem[{{Yun} {et~al.}(2001){Yun}, {Reddy}, \& {Condon}}]{yun01}
{Yun}, M.~S., {Reddy}, N.~A., \& {Condon}, J.~J. 2001, ApJ, 554, 803

\bibitem[{{Zhang} {et~al.}(2009){Zhang}, {Reiprich}, {Finoguenov}, {Hudson}, \&
  {Sarazin}}]{zha09}
{Zhang}, Y.-Y., {Reiprich}, T.~H., {Finoguenov}, A., {Hudson}, D.~S., \&
  {Sarazin}, C.~L. 2009, ApJ, 699, 1178

\end{thebibliography}

\end{document}